\documentclass[11pt,a4paper]{article}
\pdfoutput=1
\usepackage{jheppub}
\usepackage{amsfonts}
\usepackage{bbm}
\usepackage{graphicx}
\usepackage{caption}
\usepackage{subcaption}
\usepackage[normalem]{ulem}
\usepackage{amsmath,amssymb,epsfig}
\usepackage{xspace}
\usepackage{comment}
\usepackage{url}

\def\one{{\,\hbox{1\kern-.8mm l}}}
\newcommand{\Dslash}{\not{\hbox{\kern-4pt $D$}}}
\newcommand{\pdslash}{\not{\hbox{\kern-2pt $\partial$}}}

\newcommand{\Comment}[1]{{}}
\newcommand{\diag}{{\rm diag}}

\def\IZ{{\mathbb Z}}

\newcommand{\boldB}{\hbox{\boldmath $B$}}

\newcommand{\bc}{\begin{center}}
\newcommand{\ec}{\end{center}}
\newcommand{\ba}{\begin{array}}
\newcommand{\ea}{\end{array}}
\newcommand{\beq}{\begin{equation}}
\newcommand{\eeq}{\end{equation}}
\newcommand{\bea}{\begin{eqnarray}}
\newcommand{\eea}{\end{eqnarray}}
\newcommand{\bmx}{\begin{pmatrix}}
\newcommand{\emx}{\end{pmatrix}}
\newcommand{\nn}{\nonumber}

\newcommand{\be}{\begin{equation}}
\newcommand{\ee}{\end{equation}}

\newcommand{\p}{\phi}

\newcommand{\z}{\zeta}

\newcommand{\del}{\partial}
\newcommand{\half}{{\frac{1}{2}\,}}

\newcommand{\eref}[1]{Equation~(\ref{#1})}

\newcommand{\vm}{{\vec{m}}}

\newcommand{\valpha}{{\vec\alpha}}
\newcommand{\vbeta}{{\vec\beta}}

\newcommand{\vhalf}{{\vec {\frac12} }}

\newcommand{\hB}{{\widehat{B}}}
\newcommand{\hC}{{\widehat{C}}}

\newcommand{\Renyi}{R\'enyi\xspace}

\def\IC{\mathbb{C}}

\def\IZ{\mathbb{Z}}

\newcommand\sfrac[2]{{\textstyle\frac{#1}{#2}}}

\newcommand\shalf{{\textstyle\frac12}}

\normalsize

\def\a{\alpha}
\def\b{\beta}

\def\v{\varphi}

  \def\w{\omega}

\def\t{\tau}

\def\z{\zeta}

  \def\w{\omega}

\def\p{\partial}
\def\be{\begin{equation}}
\def\ee{\end{equation}}
\def\bea{\begin{eqnarray}}
\def\eea{\end{eqnarray}}
\def\ba{\begin{align}}
\def\ea{\end{align}}

\newcommand{\bem}{\begin{pmatrix}}
\newcommand{\eem}{\end{pmatrix}}

\def\={\;  = \;}
\def\+{\, + \,}

\def\wh{\widehat}
\def\bar{\overline}

\def\rt2{\sqrt{2}}

\newcommand{\nth}{$n^\text{th}$\xspace}

\title{Entanglement, Replicas, and Thetas}

\author{Sunil Mukhi$^a$,}
\author{Sameer Murthy$^b$,}
\author{and Jie-Qiang Wu$^c$}

\affiliation{$^a$ Indian Institute of Science Education and Research,\\
Homi Bhabha Rd, Pashan, Pune 411 008, India}
\affiliation{$^b$ Department of Mathematics, King's College London,\\
The Strand, London WC2R 2LS, U.K}
\affiliation{$^c$ Department of Physics and State Key Laboratory of Nuclear Physics and Technology,\\ Peking University, Beijing 100871, P.R. China}

\emailAdd{sunil.mukhi@gmail.com}
\emailAdd{sameer.murthy@kcl.ac.uk}
\emailAdd{jieqiangwu@pku.edu.cn}

\abstract{We compute the single-interval \Renyi entropy (replica partition function) for free fermions in 1+1d at
finite temperature and finite spatial size by two methods: (i)~using the higher-genus partition function on the
replica Riemann surface, and (ii)~using twist operators on the torus. We compare the two answers for a restricted
set of spin structures, leading to a non-trivial proposed equivalence between higher-genus
Siegel~$\Theta$-functions and Jacobi~$\theta$-functions. We exhibit this proposal and provide substantial
evidence for it.
The resulting expressions can be elegantly written in terms of  Jacobi forms.
Thereafter we argue that the correct \Renyi entropy for modular-invariant free-fermion theories,
such as the Ising model and the Dirac CFT, is given by the higher-genus computation summed over all spin
structures. The result satisfies the physical checks of modular covariance, the thermal entropy relation, and
Bose-Fermi equivalence.}

\begin{document}

\maketitle

\section{Introduction}

The study of entanglement measures in quantum field theories has provided a rich set of results that
illuminate these theories as well as their holographic duals (for a comprehensive review, see~\cite{Rangamani:2016dms}).
In this work we focus on some of the simplest
known quantum field theories, namely free conformal field theories in $1+1$ dimensions, and address
some puzzles that arises when one computes \Renyi and entanglement entropies for systems of finite size
and at finite temperature.

The first such computation was performed in~\cite{Azeyanagi:2007bj} in the case of $(1+1)$-dimensional fermions 
with a single entangling interval. This calculation was performed using the 
replica trick~\cite{Calabrese:2004eu,Cardy:2007mb,Calabrese:2009qy} in a standard way and computing 
two-point functions of twist operators on the torus. For fixed fermion boundary conditions (spin structure) on the 
torus, one finds a result in terms of product of Jacobi theta functions. Let us take the size of the spatial circle to be~$L$ 
and the entangling interval to be~$\ell$, and define a scaled interval $z_{12}=\frac{\ell}{L}$ that lies between~0 and~1. 
Then the entaglement entropy~$S_\text{EE}(z_{12})$ satisfies the relation~\cite{Azeyanagi:2007bj}:
%was found to satisfy the relation~\cite{Azeyanagi:2007bj}:
\be 
\lim_{z_{12}\rightarrow 0} \bigl( S_\text{EE}(1-z_{12}) -S_\text{EE}(z_{12}) \bigr)
\=S_\text{thermal} \,. 
\ee
%This means the entanglement entropy of the whole system equals the thermal entropy up to a UV cut-off term. 
%In this paper, we refer to this as the ``thermal entropy relation''. It was originally proposed in~\cite{Azeyanagi:2007bj} 
This ``thermal entropy relation'' was originally proposed in~\cite{Azeyanagi:2007bj} 
from a holographic point of view and later derived directly in CFT in~\cite{Cardy:2014jwa}, 
see also~\cite{Chen:2014unl,Chen:2014ehg,Chen:2014hta}.
The free-fermion calculation was subsequently streamlined and generalised \cite{Herzog:2013py} to the massive case,
still for a fixed spin structure.

Path-integral computations of the \Renyi entropy at finite size and temperature give an answer that is
(at least locally) analytic in the modular parameter $\tau$ of the torus and the interval size $z_{12}$. 
Thus it is natural to ask if the result is modular covariant\footnote{Physically
the length of the entangling interval is real, and the modular transformation $S$ transforms~$z_{12}$ from real
to imaginary values. Thus the modular transformation of entanglement entropy is a ``temporal'' version
of it. However, the path integral calculation is naturally performed for a generic complex interval on the
2d torus spanned by ordinary space and Euclidean time. The result is holomorphically factorised in the
interval size~$z_{12}(\bar{z}_{12})$, and it is meaningful to ask in this analytically continued setting if it is 
modular-covariant. Note that in a different context, that of multi-interval entanglement at zero
temperature, it has also been emphasised in \cite{Headrick:2012fk} that entanglement
entropy should satisfy both modular invariance and Bose-Fermi duality.}.
In this context the following puzzle was raised in \cite{Lokhande:2015zma}. Firstly it was observed
that, although the replica partition function for any fixed spin structure is not modular covariant with respect to
the full modular group of the torus, one can obtain a modular covariant answer by summing over
all four torus spin-structures for the fermions.
While this sum over torus spin-structures may seem to be a natural observable from the torus point of view,
it does not satisfy Bose-Fermi equivalence, which equates the free fermion theory (after summing over spin
structures) to a suitable free boson theory. Furthermore, this summed result does not obey the
thermal entropy relation above. 

In this paper we resolve these puzzles.
Two concrete examples, on which we will focus for the most part in this paper, are the Ising model
i.e.~one Majorana fermion summed over spin structures, and the Dirac CFT i.e.~one Dirac fermion summed
over spin structures. Each of these is a well-defined~2d CFT with a known partition function, and more
general examples can be found in \cite{Elitzur:1986ye}. The \Renyi/entanglement entropies of these theories
exhibit the above puzzles very precisely. For example the Dirac CFT is dual to a free boson theory
with~$R=1$. However, the free Dirac fermion with any fixed spin structure is not equivalent to a free boson theory.
The thermal entropy puzzle is also illustrated clearly in this system.
The twist-operator computation exhibits an ambiguity in whether the spin structure should be summed over 
before or after taking the product of~$\theta$-functions across replicas.
As shown in~\cite{Lokhande:2015zma}, neither way of performing the sum satisfies the thermal entropy relation.
A strong form of the thermal entropy relation, as explained in~\cite{Cardy:2014jwa,Chen:2014ehg}, says that
the \nth replica partition function $Z_n(z_{12},\tau)$ should reduce in the limit $z_{12}\sim 0$ to $Z_1(\tau)^n$
(up to the prefactor that encodes the singularity induced by the collision of the two twist operators),
while in the limit $z_{12}\sim 1$ it should go over to~$Z_1(n\tau)$. What was shown in~\cite{Lokhande:2015zma} 
is that the former prescription only agrees with this prediction at small intervals $z_{12}\sim 0$,
while the latter prescription only agrees with it at large intervals~$z_{12}\sim 1$.
This means that the twist-operator computation, at least in the form currently known, is inadequate to
compute the (finite size, finite temperature) \Renyi entropy of the free Majorana (Ising)/free Dirac 
CFT.
Subsequent to~\cite{Lokhande:2015zma}, issues regarding summing over fermion spin structures 
when computing \Renyi entropies have been discussed in different contexts 
in~\cite{Coser:2015eba, Coser:2015dvp, Herzog:2016ohd, Zayas:2016drv, Prudenziati:2016dbc}.

The replica partition function has also been computed at finite size and temperature for free bosons compactified 
at an arbitrary radius $R$ in the target space. The correct answer appears in~\cite{Chen:2015cna} (building on 
previous work in~\cite{Datta:2013hba}, in turn based on the orbifold twist-operator computations of \cite{Atick:1987kd}).
In \cite{Chen:2015cna} the answer was shown to satisfy the thermal entropy relation, while in~\cite{Lokhande:2015zma} 
it was shown to also be modular covariant. This result is essentially the higher-genus partition function of the free boson 
theory on a Riemann surface of genus $n$, where $n$ is the number of replicas. This Riemann surface has a fixed and 
rather special period matrix~$\Omega$ determined by two complex variables that are the physical parameters of the 
original problem: the modular parameter~$\tau$ of the original unreplicated torus and the length $z_{12}$ of the sub-system.

We propose here that the only correct way to compute the  \Renyi entropy of modular-invariant free fermionic theories at finite temperature and size is through the higher-genus relica partition function approach\footnote{Multi-interval \Renyi entropies for free fermions at zero temperature have been studied in this way in \cite{Calabrese:2010he,Coser:2013qda} and WZW models have been studied in \cite{Schnitzer:2015ira,Schnitzer:2016xaj}.}.
The full partition function is computed on the genus-$n$ replica surface as a
sum over all the $2^{2n}$ spin structures\footnote{The partition function
vanishes for odd spin structures, so only the $2^{n-1}(2^n+1)$ even spin structures contribute to this sum.} of this surface, 
$\valpha=(\alpha_1,\alpha_2,\cdots,\alpha_n)$ and $\vbeta=(\beta_1,\beta_2,\cdots,\beta_n)$
whose entries are independently chosen to be any integer or half-integer.
We show that the final answer satisfies~(i) modular covariance on the original torus and~(ii) the thermal entropy
relation. Bose-Fermi equivalence is relatively trivial because the higher-genus answer is already known to
satisfy it for any period matrix $\Omega$, and here we have simply specialised $\Omega$ to the replica surface.
Furthermore, if we focus on special spin structures of the higher genus partition function, the resulting expressions 
turn out to have interesting symmetry properties, in that they transform as Jacobi forms in the variables~$(\tau,z_{12})$.

Our proposal leads to some non-trivial checks. As we have indicated above, and will elaborate in what follows, 
for certain fixed spin structures the $n$-th R\'enyi entropy can be calculated in two distinct ways---as a partition function 
on the genus-$n$ replica surface, and as a two-point function of the \nth twist operator on the original torus.
These spin structures are the special ``diagonal'' higher-genus spin structures that obey the replica symmetry, 
namely those of the type $\valpha_\text{diag}=(\alpha,\alpha,\cdots,\alpha)$, and $\vbeta_\text{diag}=(\b,\b,\cdots,\b)$,
where~$(\alpha,\beta)$ is the spin structure on the original torus\footnote{We will often need to distinguish
the original torus from the replica surface which is an $n$-fold copy of the original torus glued pairwise along a cut.}.
The higher-genus answer is expressed in terms of Siegel~$\Theta$-functions with the special 
characteristics~$\biggl[ \begin{matrix} \valpha_\text{diag} \\ \vbeta_\text{diag} \end{matrix} \biggr]$.
The twist-operator approach involves computing two-point functions on the original torus, and
the answer is expressed in terms of Jacobi $\theta$-functions with spin structure~$(\alpha,\beta)$. We find that these
two calculations are equivalent only if a very precise identity is satisfied, which relates
Siegel~$\Theta$-constants evaluated on the special replica period matrix to a product of Jacobi $\theta$-functions
of the variables~$(\tau,z_{12})$.

We study this proposed identity in detail, showing independently that each side has the same periodicity in
the entangling interval $z_{12}$ as well as the same transformations under modular transformations of the
original torus. As a result both sides transform as Jacobi forms of the same weight and index.
The identity itself turns out to be quite non-trivial, and we check it by expanding
each side in powers of $z_{12}$. At low orders we demonstrate the identity explicitly for all~$n$.
For higher orders, we consider the special case of two replicas $n=2$, and evaluate the difference of
the two sides up to 40th order in $z_{12}$ to find an equality result in each order.
This constitutes strong evidence that the higher-genus and twist-field computations agree.

Next we elaborate on a certain interesting aspect of the results. As defined, the scaled interval length~$z_{12}$ 
takes values between~0 and~$1$. Thus, in principle the cut should lie within a fundamental region of the original torus. 
Once we analytically continue in the cut-length~$z_{12}$, however, there are other paths connecting the same 
two endpoints that wind around the two cycles of the torus. The inequivalent paths turn out to be those that 
wind~$0,1,\cdots, n-1$ times around either cycle.
The fundamental region of our problem is therefore a torus of sides~$(n,n\t)$.
We find, in accordance with this intuition, that the partition function has a
periodicity\footnote{There is a small subtlety that for even~$n$
each fixed-spin-structure partition function transforms into itself under shifts of~$2n$ and~$2n\tau$, but
in fact it transforms into a different spin-structure under shifts of~$n$ and~$n\tau$ so that the full summed
partition function transforms correctly under shifts of the lattice~$\IC/(n\tau\IZ+n\IZ)$.}
%(i.e.~up to a phase, in the sense of Jacobi forms)
under shifts of~$z_{12}$ by the lattice~$\IC/(n\tau\IZ+n\IZ)$.

The plan of the paper is as follows. In Section~\ref{sec:highergenus} we discuss the special higher-genus surfaces
arising from the replica trick and study their partition functions. We then present our proposal of equality of the two ways 
of computing these partition functions---the direct higher-genus method and the twist-operator method. 
In Section~\ref{sec:checks} we discuss various checks of our proposal for arbitrary~$n$, including the symmetry 
properties and the small-interval expansion.
In Section~\ref{sec:nequals2} we focus on the~$n=2$ case and verify our proposed relation to high order in
an expansion in~$z_{12}$. In Section~\ref{sec:sumspin} we sum over spin structures and show that the
answer obeys the various physical properties that we expect. In Section~\ref{sec:summary} we summarize our results and 
conclude. In the two appendices we discuss, respectively, the periodicity and modular properties of the higher-genus
and twist-operator expressions.

\section{Higher-genus surfaces and replica partition functions \label{sec:highergenus}}

\subsection{Period matrix of the replica higher-genus surface}

Let us start by summarising the methodology of computation of the replica partition function for a CFT at finite size and 
temperature, i.e.~on the torus. As is well-known, in such a case the replica trick gives rise to a Riemann surface made 
up by joining $n$ copies of the torus (where $n$ is the number of replicas) sequentially along the entangling interval. 
The result is a genus-$n$ surface, though a very special one. By definition, the replica partition function should be simply 
the partition function of the given CFT on this genus-$n$ surface. On the other hand, one can introduce ``twist operators'' 
whose non-local OPE's with the free fermions have the effect of ``transporting'' them from one replica to the next. The 
replica partition function is then identified with the correlation function of these twist operators. Next, a diagonalisation procedure,
described for example in~\cite{Calabrese:2009qy}, reduces the problem to the correlation function of twist fields on 
a \emph{single replica surface}. In this approach the computation is performed on a torus rather than on the replica surface.

The question we wish to first address is whether the two computations, which we call the  ``higher-genus calculation'' 
and the ``twist-field calculation'', are equivalent at least for a fixed free-fermion spin structure. Since there are more spin structures 
in genus-$n$ than on the torus, this question is not well-defined. Nevertheless, one can ask if the 
higher-genus calculation for the special class of spin structures of the form $\valpha_\text{diag}=(\alpha,\alpha,\cdots,\alpha)$ (and 
similarly for $\vbeta$) gives the same answer as the twist-operator calculation for a given spin structure $(\alpha,\beta)$ on the torus. 
Note that the two calculations are quite distinct. The higher-genus one uses general results about free-fermion partition functions 
on higher-genus surfaces \cite{AlvarezGaume:1986es,Dijkgraaf:1987vp}. One takes the appropriate result and inserts the period 
matrix of the replica Riemann surface, which as we will soon see is quite special. 
On the other hand the twist-operator computation uses the torus correlation function of
twist operators. With the higher-genus spin structures restricted as above, we will find a precise equivalence
between them. This result follows from a nontrivial identity between a genus-$n$ Siegel~$\Theta$-constant
evaluated for a specific subclass of period matrices, and a product of genus-1 Jacobi~$\theta$-functions.
We will show that both sides of the identity are in fact Jacobi forms \cite{eichler1985the} on an $n$-fold/$2n$-fold cover of the
original torus (depending on whether~$n$ is odd or even).

Riemann surfaces of genus $n$ are parametrised by their period matrix
$\Omega$ which is in general a complex symmetric $n\times n$ matrix. As is
well-known, this has more parameters than necessary to describe a general
Riemann surface of this genus. The problem of determining which subset
of $\Omega$'s correspond to a Riemann surface is called the Schottky
problem. However the situation of interest to us is much simpler, as we are only
interested in the special Riemann surfaces that arise via the replica
trick. The surface defined by taking $n$ replicas of a
torus with a single interval has only two independent
moduli,~$z$ and~$\tau$, where $\tau=i\frac{\beta}{L}$ is the modular
parameter of the original torus and, as already indicated, $z_{12}=\frac{\ell}{L}$ is the relative
length of the entangling interval. Thus, we would like to express $\Omega_{ij}$ as a function
of~$z_{12}$ and~$\tau$.

The desired answer follows from the constructions of cut differentials for the Riemann surface of
interest \cite{Atick:1987kd,Datta:2013hba,Chen:2015cna}. These are given, for example, in the Appendix
of \cite{Chen:2015cna}\footnote{The arguments of our cut differentials are shifted with respect to those
of~\cite{Chen:2015cna} so that $\w_\text{here}(z)=\w_\text{there}\big(z+(1-\frac{k}{n})z_1+\frac{k}{n}z_2\big)$.
This does not, of course, affect the periods.}:
\be
\omega_k(z,z_{12},\tau) \,:= \, \frac{\theta_1(z|\tau)}{
\theta_1\Big(z+\frac{k}{n}z_{12}\Big|\tau\Big)^{1-\frac{k}{n}}\theta_1\Big(z-(1-\frac{k}{n})z_{12}
\Big|\tau\Big)^{\frac{k}{n}}} \,,
\label{cutdiff}
\ee
where $n$ is the number of replicas which equals to the genus of Riemann surface and $k=0,1,2,\cdots,n-1$ labels the linear independent differential cuts. Selecting a basis of cycles $A_a,B_a$, we have:
\be
A_{ak}\=\int_{A_a}\omega_k,\qquad B_{ak}\=\int_{B_a}\omega_k \,.
\ee
The period matrix of the Riemann surface is then $\Omega=B\cdot A^{-1}$.

So far we have not specified a basis of cycles. However there is something special about the genus-$n$ Riemann 
surface obtained by gluing $n$ tori sequentially along a cut, namely the fact that the gluing procedure does not introduce 
any new handles to the surface. It simply connects~$n$ genus-1 surfaces into a single genus-$n$ surface. This is in contrast 
to the surfaces obtained by gluing several complex planes along a pair of cuts~\cite{Calabrese:2009ez}, relevant to the case  
where the entangling region is made of two disjoint components (at zero temperature and infinite spatial size). 
In the latter case it is in fact the gluing procedure that introduces the handles to the resulting surface. Therefore in that case, 
any basis of cycles on the replica Riemann surface must necessarily involve the cuts. For our case, each component torus of 
the replica surface already has a pair of canonical cycles, the usual $(A,B)$ pair for a torus. These continue to be valid cycles 
of the glued surface of genus~$n$, and it turns out very convenient to choose them as a basis for the latter. Accordingly, 
from now on $A_a,B_a$ will be taken to be the cycles of the $a$-th torus component of the glued replica Riemann surface. 
This choice will considerably simplify the analysis of the problem.

By setting $k=0$, we see that $A_{00}=1, B_{00}=\tau$. We can now relate all the entries $A_{ak}$ to~$A_{0k}$, and 
likewise $B_{ak}$ to $B_{0k}$. For this, notice that $\omega_k$ picks up a phase $\alpha^k$ where~$\alpha=e^{\frac{2\pi i}{n}}$, 
when we go from one replica to the next one above it. It follows that:
\be
A_{ak}\=\alpha^{ak}A_{0k}\,,\qquad B_{ak}\=\alpha^{ak}B_{0k} \,,
\ee
where:
\be
A_{0k}(z_{12},\tau)\=\int_0^1\omega_k(z,z_{12},\tau)\,dz\,,\qquad
B_{0k}(z_{12},\tau)\=\int_0^\tau\omega_k(z,z_{12},\tau)\,dz \,.
\label{ABdef}
\ee

We can think of the above as products of the matrix $M$ with entries $M_{ak}=\alpha^{ak}$ with the diagonal matrix 
$A^D=\diag(A_{0k})$ or $B^D=\diag(B_{0k})$:
\be
A\=MA^D\,,\qquad B\=MB^D \,.
\ee
Then we have:
\be
\Omega\=MB^D(A^D)^{-1}M^{-1}\=MC M^{-1} \,,
\ee
where we have defined the diagonal matrix $C=\diag(C_k)$ with components\footnote{In the notation 
of~\cite{Chen:2015cna}, this corresponds to $C_k=\frac{W_2^2(k)}{W_1^1(k)}$.}:
\be
C_k\;\equiv\; \frac{B_{0k}}{A_{0k}} \,.
\label{Ckdef}
\ee
From the discussion above, we know that $C_{0}=\tau$.
The inverse of $M(\alpha)$ is $\frac{1}{n}M(\alpha^{-1})$ and one has:
\be
\Omega_{ab}\=\frac{1}{n}\sum_{k=0}^{n-1}\alpha^{(a-b)k}C_k \,.
\label{omegaun}
\ee
Using the property that $C_k=C_{n-k}$, which is easily verified, we can check that the above matrix is symmetric 
as it should be. Hence it can equivalently be written as:
\be
\Omega_{ab}\=\frac{1}{n} \sum_{k=0}^{n-1}\cos\left(\frac{2\pi(a-b)k}{n}\right)C_{k} \,.
\label{omegadef}
\ee
As examples, for $n=2,3$ we have:
\be
\begin{split}
n=2: & \quad \Omega\=\half \begin{pmatrix} \tau+C_1 & \tau-C_1\\
\tau-C_1 & \tau+C_1\end{pmatrix} \,,\\
 n=3: & \quad \Omega\=\frac13 \begin{pmatrix} \tau+C_1+C_2 & \tau-\half(C_1+C_2)& \tau-\half(C_1+C_2)\\
 \tau-\half(C_1+C_2) & \tau+C_1+C_2 & \tau-\half(C_1+C_2)\\
  \tau-\half(C_1+C_2) & \tau-\half(C_1+C_2) & \tau+C_1+C_2
 \end{pmatrix} \,.
\end{split}
\label{Omegaexamples}
\ee
For every~$n$ the period matrix satisfies the identity that the sum of all elements in a row (or column)
is equal to $\tau$:
\be
\sum_b\Omega_{ab}\=\tau \,.
\label{Ominv}
\ee
For future use, we note here some additional identities that it satisfies:
\be
\begin{split}
\det\Omega&\=\prod_{k=0}^{n-1}C_k \,, \\
\Omega^{-1}_{ab}&\=\frac{1}{n}\sum_{k=0}^{n-1}\alpha^{(a-b)k}\frac{1}{C_k} \,.
\end{split}
\label{Ominv2}
\ee

\subsection{Relationship between higher-genus partition function and twist-operator computation}

We would now like to describe the relationship between the two ways of computing the
replica partition function: one as a higher-genus partition function evaluated on the restricted family of
period matrices defined above, and the other as a correlation function of twist operators on the torus.

For the former, we first write the genus-$n$ Siegel $\Theta$-function:
\be
\Theta^{(n)}(0|\Omega)\bigg[\begin{matrix}\valpha \\ \vbeta\end{matrix}\bigg]
\; := \;\sum_{{\vec m}\in \IZ^n} \exp\Big(\pi i \, ({\vec m}+\valpha)\cdot \Omega\cdot ({\vec m}+\valpha)
+2\pi i \, ({\vec m}+\valpha)\cdot\vbeta\Big) \,,
\ee
where the characteristics $\valpha=(\alpha_1,\alpha_2,\cdots,\alpha_n)$ and
$\vbeta=(\beta_1,\beta_2,\cdots,\beta_n)$ are two $n$-component vectors whose entries are independently
chosen to be any integer or half-integer. The independent choices are $0,\half$, and all other choices can be
related to these. This theta-function is one of the factors in the free-fermion partition function in higher-genus.
In the context of free fermions, the characteristics arise as boundary conditions along different cycles of the
Riemann surface, namely spin structures.

The genus-$n$ fermion partition function for a single Dirac fermion with arbitrary higher-genus spin
structure~$\biggl[\begin{matrix}\valpha \\ \vbeta\end{matrix}\biggr]$ is \cite{AlvarezGaume:1986es,Dijkgraaf:1987vp}:
\be
Z^{(n)}_\text{Dirac}\biggl[\begin{matrix}\valpha \\ \vbeta\end{matrix}\biggr](\Omega)
\=|{\cal C}|^2 \,
\bigg|\Theta^{(n)}(0|\Omega)\bigg[\begin{matrix}\valpha \\ \vbeta\end{matrix}\bigg]\bigg|^2 \,,
\label{Dirac}
\ee
while for a single Majorana fermion, it is:
\be
Z^{(n)}_\text{Majorana}\biggl[\begin{matrix}\valpha \\ \vbeta\end{matrix}\biggr](\Omega)
\=|{\cal C}| \,
\bigg|\Theta^{(n)}(0|\Omega)\bigg[\begin{matrix}\valpha \\ \vbeta\end{matrix}\bigg]\bigg| \,,
\label{Majorana}
\ee
where ${\cal C}$ is a spin-structure-independent factor related to the determinant of an (anti-) holomorphic 
differential operator on the surface (see Eq.(5.13) of~\cite{AlvarezGaume:1986es})\footnote{As pointed out 
in~\cite{AlvarezGaume:1986es} this quantity depends not only on the moduli but on the metric of the surface
because of conformal and diffeomorphism anomalies. For us the precise metric is determined by the replica 
construction to be flat everywhere except at the end points of the cuts.}. 
The full Dirac or Majorana partition 
function is a sum over all~$2^{2n}$ spin structures arising on the higher-genus
Riemann surface:
\be\label{ZFullDirMaj}
Z^{(n)}_\text{Dirac/Majorana}(\Omega) \=\frac{1}{2^n}
\sum_{\valpha,\vbeta} Z^{(n)}_\text{Dirac/Majorana}\biggl[\begin{matrix}\valpha \\ \vbeta\end{matrix}\biggr](\Omega) \,.
\ee

We see that besides the $\Theta$-function, the replica partition function has an additional factor of a power 
of~$|{\cal C}|$. To compute it for our case, we use the fact that this determinant is independent of spin structures,
and is the same one that appears in the bosonic partition function. Indeed, it was proved long 
ago~\cite{AlvarezGaume:1986es,Dijkgraaf:1987vp} that Bose-Fermi equivalence holds on arbitrary Riemann surfaces. 
Now the free boson replica partition function~\cite{Datta:2013hba,Chen:2015cna} is expressed as a higher-genus 
boson partition function evaluated on the period matrix~$\Omega$ of the previous section. It depends on the 
radius~$R$ at which the boson is compactified. For $R=1$ one can perform standard manipulations to reduce the 
free boson partition function to a sum over $\Theta$-functions with characteristics~$\valpha,\vbeta$ times some other terms.
The result is:
\be
Z^{(n)}_\text{boson}\=
\left|\frac{\theta_1'(0|\tau)}{\theta_1(z_{12}|\tau)}\right|^{\frac16(n-\frac{1}{n})}\frac{1}{|\eta(\tau)|^{2n}}
\frac{\sum_{\valpha,\vbeta}
\Big|\Theta\bigg[\begin{matrix}\valpha \\ \vbeta\end{matrix} \bigg](0|\Omega)\Big|^2}{\prod_{k=0}^{n-1}|A_{0k}|} \,.
\label{Bosongenusn}
\ee
This bosonic partition function can be understood as arising from a classical part and a quantum part. 
The classical part is simply the higher-genus theta function, so the quantum part can be identified with the 
factor~$\frac{1}{2^g}|{\cal C}|^2$. We can thus read off this
factor to be:
\be \label{detdel0}
\frac{1}{2^g}|{\cal C}|^2 \=
\left|\frac{\theta_1'(0|\tau)}{\theta_1(z_{12}|\tau)}\right|^{\frac16(n-\frac{1}{n})}\frac{1}{|\eta(\tau)|^{2n}
\prod_{k=0}^{n-1}|A_{0k}|} \,.
\ee
Comparing Equations~\eqref{Dirac},~\eqref{ZFullDirMaj}, \eqref{detdel0}
to \eref{Bosongenusn} we see that this identification for~$|{\cal C}|^2$ implies Bose-Fermi equivalence 
for the replica higher-genus surfaces as desired. 

Note that we used the twist-operator calculation of the bosonic partition function and combined it
with the intuition that the classical paths on the higher-genus surface sum up to form the appropriate
$\Theta$-function, to deduce the determinant factor as a quotient of the two expressions.
It would be nice to check~\eref{detdel0} from a direct calculation of ${\cal C}$ as the determinant of a 
differential operator on the higher-genus surface under consideration here.

Now we examine how the higher-genus computation of the replica partition function is related to the twist-operator calculation
in~\cite{Azeyanagi:2007bj,Herzog:2013py,Lokhande:2015zma} which was carried out by computing two-point functions of
twist operators on a single torus. The latter result, for a fixed fermion spin structure $\alpha,\beta$ and for the Dirac fermion, 
is as follows:
\be
Z^{\rm twist~field}_{\alpha,\beta}\,=\,\left|\frac{\theta_1'(0|\tau)}{\theta_1(z_{12}|\tau)}\right|^{\frac16(n-\frac{1}{n})}
\prod_{k=-{\frac{n-1}{2}}}^{\frac{n-1}{2}}\left|\rule{0cm}{10mm}\right.
\frac{\theta\bigg[\begin{matrix}\alpha\\ \beta\end{matrix}\bigg]\left(\frac{k}{n}z_{12}\Big|\tau\right)}{\eta(\tau)}
\left.\rule{0cm}{10mm}\right|^2 \,.
\label{zcltorus}
\ee
For a Majorana fermion one simply takes the square root of the above expression.

The question is now whether the higher-genus approach leading to \eref{Dirac}, \eqref{Majorana} and the torus
twist-operator approach leading to \eref{zcltorus} are equivalent.
As it stands, however,  this is not a well-posed question because the former depends on $2^{2n}$
spin structures while the latter has just 4.
Thus we restrict the higher-genus expression to a fixed spin structure
$\vec{\alpha}_\text{diag}=(\alpha,\alpha,\cdots,\alpha)$ and $\vbeta_\text{diag}=(\b,\b,\cdots,\b)$.
These are very special spin structures that are taken to be the same over each ``handle'' of the genus-$g$
surface. They are certainly not the only ones that contribute to the full modular-invariant higher-genus partition
function. But they have a simple representation as a replicated partition function of one of the 4 torus spin
structures. We postpone investigation of the appearance of the other non-replica spin-structures from the
torus point of view to the future. Here we ask the question, for this for spin structure, whether the calculation 
from higher genus partition function and the calculation from twist operators are equal to each other.

Both the objects in question are squares of locally analytic expressions. Thus we take the holomorphic
square root $\chi$, so that $Z=|\chi|^2$, and examine whether the two expressions
\be
\chi_\text{higher-genus}\=
\left(\frac{\theta_1'(0|\tau)}{\theta_1(z_{12}|\tau)}\right)^{\frac{1}{12}(n-\frac{1}{n})}
\frac{1}{\eta(\tau)^n}
\frac{\Theta\bigg[\begin{matrix}\valpha_\text{diag}\\ \vbeta_\text{diag}\end{matrix}\bigg](0|\Omega)}{\sqrt{\prod_{k=1}^{n-1}(A_{0k})}}
\label{chihg}
\ee
with~$\Omega$ defined in \eref{omegadef}, and
\be
\chi_\text{twist-field} \=
\left(\frac{\theta_1'(0|\tau)}{\theta_1(z_{12}|\tau)}\right)^{\frac{1}{12}(n-\frac{1}{n})}\prod_{k=-{\frac{n-1}{2}}}^{\frac{n-1}{2}}\frac{\theta\bigg[\begin{matrix}\alpha\\ \beta\end{matrix}\bigg]\left(\frac{k}{n}z_{12}\Big|\tau\right)}{\eta(\tau)} \,,
\label{chitf}
\ee
are equal to each other.

Cancelling out common factors, we can restate the above considerations in terms of the higher-genus expression
\be
\chi_{g}(z_{12},\tau;\alpha, \beta)
\, := \,
\frac{\Theta\bigg[\begin{matrix}\valpha_\text{diag}\\ \vbeta_\text{diag}\end{matrix}\bigg](0|\Omega)}{\sqrt{\prod_{k=1}^{n-1}(A_{0k})}}\,,
\label{Ldef}
\ee
and the twist-field expression
\be
\chi_{t}(z_{12},\tau;\alpha, \beta)
\, := \,
\prod_{k=-{\frac{n-1}{2}}}^{\frac{n-1}{2}}\theta\bigg[\begin{matrix}\alpha\\ \beta\end{matrix}\bigg]\left(\frac{k}{n} z_{12}\Big|\tau\right)\,.
\label{Rdef}
\ee
Since the two expressions~\eqref{chihg} and~\eqref{chitf} come from two different ways of computing the same physical quantity, 
namely the \nth \Renyi entropy, we propose the equality:
\be
\chi_{g}(z_{12},\tau;\alpha, \beta) \= \chi_{t}(z_{12},\tau;\alpha, \beta)  \,.
\label{question}
\ee

\section{Some checks of the higher-genus partition function/twist-operator equivalence \label{sec:checks}}

\begin{figure}
\begin{center}
\includegraphics[height=5cm]{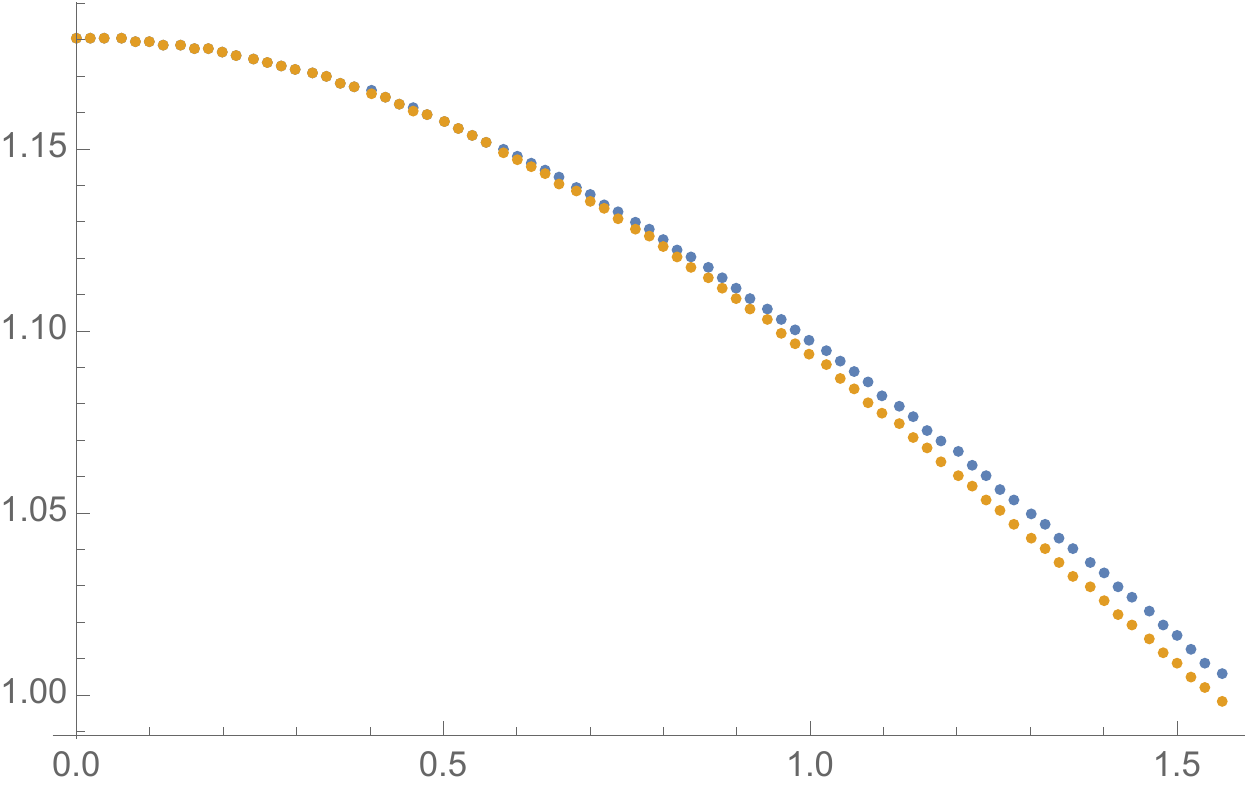}
\end{center}
\caption{Plot of the LHS of \eref{question} (yellow) and RHS of the same equation (blue) as a function of $2\pi z$ where $z$ is the interval length normalised to unity. The range is $0<2\pi z <\pi$ (i.e. $0<z<\half$). This is for Dirac fermions (for Majorana fermions one would take the square root of both sides).}
\label{questionplot}
\end{figure}

In what follows, we provide evidence for \eref{question} which is a nontrivial equality between two well-defined mathematical functions. To start with, we plot them numerically for $n=2$. We fix the modular parameter to be $\tau=i$ and the spin structure to be $(\alpha,\beta)=(0,0)$, and let $z$ take real values from 0 to $\half$.  The plots are shown in Figure~\ref{questionplot}. The proposed equality is exact, as our subsequent mathematical analysis will show, hence the slight deviations visible near $z=1$ are attributable to an inaccuracy of the numerical plot rather than of the equality itself. This in turn arises from the fact that the cut differential involves a square root and one has to choose the integration contour to avoid the branch cut. The numerical programme becomes inaccurate as the contour approaches one edge of the cut. 

Our main focus in the rest of this paper is to provide mathematical evidence for the equality using known properties of the functions on both sides. There are some basic checks that must be satisfied if it
is to hold. A zeroth check is that both sides are manifestly equal for $n=1$.
Next, when there is no cut, the genus~$n$ surface factorizes into~$n$ identical genus-1 surfaces.
In accordance with this, when~$z_{12}=0$ we have $A_{0k}=1, B_{0k}=\tau$ and $\Omega_{ab}=\tau\delta_{ab}$,
so that
\be
\chi_{g}(z_{12}=0,\tau;\alpha,\beta)\=
\biggr(\theta\bigg[\begin{matrix}\alpha\\ \beta\end{matrix}\bigg](0|\tau)\biggr)^n
\= \chi_{t}(z_{12}=0,\tau;\alpha,\beta) \,.
\ee
Now note that if $n$ is odd then $\chi_{t}$ is zero for the spin structure $\bigg[\begin{matrix}\half \\ \half \end{matrix}\bigg]$
because this is an odd spin structure on the torus and the $k=0$ term in the product therefore vanishes.
At the same time, $\chi_{g}$ vanishes because for odd genus this corresponds to an odd spin structure.
However for even $n$ there is no $k=0$ term in $\chi_{g}$ and it does not have to vanish even
for the spin structure $\bigg[\begin{matrix}\half \\ \half \end{matrix}\bigg]$. And from the higher-genus point of
view, this spin structure is now even and therefore $\chi_{t}$ also does not have to vanish.
Thus some simple checks are satisfied.

\subsection{Symmetry properties: periodicity and modularity}

The first thing to verify is that both~$\chi_{g}$ and~$\chi_{t}$ have the same periodicities and modular
transformation properties.
We start with the periodicity. It is easy to convince oneself that neither side is periodic under
$z_{12}\to z_{12}+1,z_{12}\to z_{12}+\tau$. In fact they become (quasi)-periodic only after multiple shifts.
Let us first examine the periodicities for odd $n$, and thereafter summarise the results for even $n$ which
are similar but differ slightly in some details. For odd $n$, the functions~$\chi_{g}$ and~$\chi_{t}$ are (quasi)-periodic under
\be
z_{12}\to z_{12}+n,\quad z_{12}\to z_{12}+n\tau \,.
\ee
As we show in Appendix A, the function~$\chi=\chi_{g}$ as well as $\chi=\chi_{t}$ obeys:
\be \label{periodn}
\begin{split}
\chi(z_{12}+n,\tau;\alpha,\beta)&\= \chi(z_{12},\tau;\alpha,\beta) \,, \\
\chi(z_{12}+n\tau,\tau;\alpha,\beta)&\= e^{-i\pi\frac{n(n^2-1)}{12}\tau} \,e^{-i\pi\frac{n^2-1}{6} z_{12}} \,
\chi(z_{12},\tau;\alpha,\beta) \,.
\end{split}
\ee
We see that the periodicities~\eqref{periodn} are not those of the torus but of its $n$-fold cover. As explained in
the introduction, this is due to the fact that there are non-trivial paths for the cuts connecting the two
end-points~$z_{1}$ and~$z_{2}$ which wind around the two cycles of the torus. A path that winds
around~$n$ times around either cycle of the original torus is equivalent to a closed cycle on the higher-genus
surface, and therefore the inequivalent paths are those that wind~$0,1,\cdots, n-1$ times around either cycle.
In the context of the Ising model these winding paths can be identified with disorder
operators~\cite{DiFrancesco:1997nk}. 

It is convenient to re-define the variables so that the periodicities are again those of the torus.
Defining~$Z=z_{12}/n$, and defining
\be
F(Z,\tau;\alpha,\beta) \; := \; \chi(z_{12},\tau;\alpha,\beta)  \,,
\ee
the periodicities become, for both~$\chi_{g}$ and~$\chi_{t}$:
\be
\begin{split}
F(Z+1,\tau;\alpha, \beta) &\= F(Z,\tau;\alpha, \beta) \,, \\
F(Z+\tau,\tau;\alpha, \beta) &\= e^{-2\pi i\frac{n}{12}(n^2-1)Z}e^{-\pi i\tau \frac{n}{12}(n^2-1)}\,
F(Z,\tau;\alpha, \beta) \,.
\end{split}
\label{Fper}
\ee

Next we turn to modular transformations.
In Appendix B we study the modular transformations of~$\chi_{g}(z_{12},\tau+1;\alpha,\beta)$ and~$\chi_{t}(z_{12},\tau+1;\alpha,\beta)$.
Indeed they both have the same modular behavior which can be written in terms of
the function~$F(Z,\tau+1;\alpha,\beta)$ above as:
\be
\begin{split}
F(Z,\tau+1;\alpha,\beta)
&\=e^{-i\pi n\alpha(\alpha+1)}\,F(Z,\tau;\alpha,\alpha+\beta+\shalf) \, , \\
F\left(\frac{Z}{\tau},-\frac{1}{\tau};\alpha, \beta\right)
&\=(-i\tau)^{\frac{n}{2}}e^{\frac{\pi i}{\tau}\frac{1}{12}n(n^2-1)Z^2}e^{2\pi i\alpha\beta n}F(Z,\tau; \beta, -\alpha) \,.
\label{Fmod}
\end{split}
\ee

These properties indicate that the quantities $\chi_{g}$ and $\chi_{t}$ transform as  Jacobi forms.
We recall that a Jacobi form~$\v_{k,m}(Z,\tau)$ of weight~$k$ and index~$m$  \cite{eichler1985the} obeys:
\be
\begin{split}
\v_{k,m}\left(\frac{Z}{c\tau+d},\frac{a\tau +b}{c\tau +d}\right)&\= (c\tau+d)^k e^{\frac{2\pi i mcZ^2}{c\tau+d}}
\v_{k,m}(Z,\tau) \, , \\
\v_{k,m}(Z+\mu+\lambda\tau)&\=e^{-2\pi im(\lambda^2\tau+2\lambda z)}\v_{k,m}(Z,\tau) \,.
\end{split}
\ee
We see that our quantity $F(Z,\tau)$ satisfies these equations, upto a phase and also up to a change in the spin
structure $(\alpha,\beta)$\footnote{Both of these can be eliminated by going to a suitable congruence subgroup of SL(2,Z), but we will not write out the details of this here.}.
From the transformation properties of~Equations~\eqref{Fper} and~(\ref{Fmod}) we deduce
that~$\chi_{g}$ and~$\chi_{t}$ have weight and index:
\be
k\=\frac{n}{2}\,,\qquad m\=\frac{n(n^2-1)}{24} \,.
\ee

For even~$n$, the above discussion has to be slightly modified. Under shifts of~$n$ and~$n\tau$, both~$\chi_g$ 
and~$\chi_t$ change their spin-structure (in the same way). Therefore to find their quasi-periodicity at fixed 
spin-structure, we must double the shifts. Indeed we show in Appendix~A that, for even~$n$, $\chi_g$ and~$\chi_t$ 
are both periodic under $z_{12}\to z_{12}+2n$ and quasi-periodic with the same prefactor under $z_{12}\to z_{12}+2n\tau$.

We note here that the function~$\chi_{t}$ is manifestly holomorphic in~$z_{12}$, which allows us to write a
double Fourier expansion:
\be
\chi_{t}(Z,\t;\a,\b) \= \sum_{n,r} c(n,r) \, q^{n} \, \z^{r} \,.
\ee
From the properties of Jacobi theta functions we also see that~$n$ only takes positive values in some
one-dimensional lattice depending on~$n$ and the spin structure (i.e.~$n,r$ need not be whole integers).
Combined this observation with the transformation
properties above, we see that~$\chi_{t}$ is really a weak Jacobi form in the sense of~\cite{eichler1985the}.

\subsection{Small-interval expansion}

A stronger check is to expand both sides in powers of the interval size $z_{12}$.
We have already seen that~$\chi_{g}$ and~$\chi_{t}$ agree at~$z_{12}=0$. As we see below, the coefficient
of~$z_{12}^{2}$ vanishes on both sides, and the first non-zero coefficient is at~$O(z_{12}^{4})$.
In this subsection we consider both $n$ (the number of replicas) and the spin structure to be arbitrary
(and correspondingly we suppress the spin structure label). We compare the expressions $\chi_{g}$ and $\chi_{t}$
to order~$O(z_{12}^{4})$ in the interval size, and find non-trivial agreement for all~$n$.

On expanding $\chi_{t}$ from \eref{Rdef}, we get:
\be
\begin{split}
\chi_{t}(z_{12},\tau)&\=\prod_{k=-\frac{n-1}{2}}^{\frac{n-1}{2}}\theta\Bigl(\frac{k}{n}z_{12}\Big|\tau\Bigr) \,, \\
& \= \theta(0|\tau)^n\prod_{k=-\frac{n-1}{2}}^{\frac{n-1}{2}}\left(1+\frac{kz_{12}}{n}\frac{\theta'}{\theta}
+\half \frac{k^2z_{12}^2}{n^2} \frac{\theta''}{\theta}+{\cal O}(z_{12}^3)\right) \,,
\end{split}
\ee
where~$'$ denotes the derivative with respect to the first argument.
Here and in the following, for simplicity of notation, whenever the first argument of the $\theta$ function
is suppressed it is understood to be 0. Since each Jacobi $\theta$-function is either even or odd as a function
of its first argument, alternate terms in the above expansion vanish---though we will carry along all terms in
the interest of a uniform notation. For the odd spin structure there are also vanishing $\theta(0|\tau)$ factors
in denominators and in front of the full expression, of course the two cancel each other out. In this way we can
use the same formulae for all spin structures.

Let us now switch to the variable $Z=z_{12}/n$ defined earlier.  Working to quadratic order in $Z$, the last factor above is:
\bea
&&\prod_{k=-\frac{n-1}{2}}^{\frac{n-1}{2}}\biggl(1+kZ\frac{\theta'}{\theta}+\half k^2 Z^2 \frac{\theta''}{\theta}
\biggr) \,,  \nn\\
&&\qquad \qquad \qquad \= 1+ Z^2\biggl(\frac{\theta'}{\theta}\biggr)^2\sum_{k_1\ne k_2=-\frac{n-1}{2}}^{\frac{n-1}{2}}k_1k_2 +\half Z^2 \frac{\theta''}{\theta}\sum_{-\frac{n-1}{2}}^{\frac{n-1}{2}}k^2 + {\cal O}(Z^3) \,, \nn\\
&& \qquad \qquad  \qquad \=1+ Z^2\biggl(\frac{\theta''}{\theta}-\Big(\frac{\theta'}{\theta}\Big)^2\biggr)\sum_{k=1}^{\frac{n-1}{2}}k^2 \,, \nn\\
&&\qquad \qquad  \qquad \=1+ Z^2\,\frac{n(n^2-1)}{24}
\biggl(\frac{\theta''}{\theta}-\Big(\frac{\theta'}{\theta}\Big)^2\biggr) \,.
\label{genusonecorr}
\eea
Note that the term of order $Z$ vanishes.

Thus we have:
\be
\chi_{t}(z_{12},\tau) \= \theta(0|\tau)^n\left(1+ Z^2\,\frac{n(n^2-1)}{24}\Big(\frac{\theta''}{\theta}-
\Big(\frac{\theta'}{\theta}\Big)^2\Big)+{\cal O}(Z^4)\right) \,.
\ee

Now we would like to compare $\chi_{t}$ with $\chi_{g}$ defined in \eref{Ldef} for general $n$ and general spin
structures, up to ${\cal O}(Z^2)$. For this, let us recall the cut differential and express it in terms of $Z$:
\be
\omega_k\=\frac{\theta_1(z)}{\theta_1(z+kZ)^{(1-\frac{k}{n})} \, \theta_1(z-(n-k)Z)^{\frac{k}{n}}} \,,
\ee
where the $\tau$-dependence has been suppressed to simplify the notation. Notice that this is invariant under the simultaneous transformation $k\to n-k$ and $Z\to -Z$.
We expand this to second order in $Z$ and find:
\be
\omega_k(z)\=1+\half k(k-n)\bigl(\log\theta_1(z)\bigr)'' Z^2+{\cal O}(Z^3) \,.
\ee
Again, the term of order $Z$ vanishes. The next step is to compute the integrals:
\be
\begin{split}
A_{0k}\; := \; \int_0^1 \omega_k\, dz&\=1+\half k(k-n) Z^2\int_0^1
\bigl(\log\theta_1(z)\bigr)''+{\cal O}(Z^3) \,, \\
&\= 1+{\cal O}(Z^3) \,, \\
B_{0k} \; := \; \int_0^\tau \omega_k\, dz&\=\tau+\half k(k-n) Z^2\int_0^\tau
\bigl(\log\theta_1(z)\bigr)''+{\cal O}(Z^3) \,, \\
&\= \tau-i\pi k(k-n) Z^2+{\cal O}(Z^3) \,.
\end{split}
\ee
Here we used the identities:
\be
\begin{split}
\bigl(\log\theta_1\bigr)'(z+1)&\=\bigl(\log\theta_1\bigr)'(z) \,,\\
\bigl(\log\theta_1\bigr)'(z+\tau)&\=\bigl(\log\theta_1\bigr)'(z)-2\pi i \,.
\end{split}
\ee
It follows that:
\be
C_k \; :=\; \frac{B_{0k}}{A_{0k}}\=\tau-i\pi k(k-n) Z^2+{\cal O}(Z^3) \,.
\ee

Next we compute the matrix $\Omega$ from this using \eref{omegaun}:
\be
\begin{split}
\Omega_{ab}&\=\frac{1}{n}\sum_{k=0}^{n-1}\cos\left(\frac{2\pi(a-b)k}{n}\right)C_k \,,\\
&\=\tau\delta_{ab}-\frac{i\pi}{n} Z^2\sum_{k=0}^{n-1}k(k-n)\cos\left(\frac{2\pi(a-b)k}{n}\right)+\cdots \,.
\end{split}
\ee
Defining:
\be
f(a-b) \= -\frac{i\pi}{n}\sum_{k=0}^{n-1}k(k-n)\cos\left(\frac{2\pi(a-b)k}{n}\right) \,,
\ee
we may write:
\be
\Omega_{ab}\=\tau\delta_{ab}+Z^2f(a-b)+{\cal O}(Z^3) \,.
\ee

Now we are in a position to evaluate the function $\chi_{g}$ in \eref{Ldef} to second order in~$z_{12}$.
To this order, we have seen above that~$A_{0k}=0$ for all~$k$. Thus to this order, we have:
\be
\chi_{g}(z_{12},\tau;\alpha,\beta)\=\Theta\bigg[\begin{matrix}\valpha_\text{diag}\\ \vbeta_\text{diag}\end{matrix}\bigg]
\Big(0\Big|\Omega(z_{12},\tau)\Big) \,,
\ee
where we recall that $\valpha_\text{diag}:=(\alpha,\alpha,\cdots,\alpha)$ and similarly for $\vbeta$.

Expanding the $\Theta$ function we get:
\be
\begin{split}
&\Theta\bigg[\begin{matrix}\valpha_\text{diag}\\ \vbeta_\text{diag}\end{matrix}\bigg]
\Big(0\Big|\Omega(z_{12},\tau)\Big)\=\sum_{\{m_a\}}\exp\left\{i\pi \sum_{a=1}^n (m_a+\alpha)^2\tau+2\pi i (m_a+\alpha)\beta\right\}\times\\
&\qquad\qquad\qquad\qquad\qquad\qquad \exp\left\{i\pi \sum_{a,b=1}^n (m_a+\alpha)(m_b+\alpha)f(a-b)z_{12}^2\right\}\\
&\qquad\qquad\qquad\qquad\=\biggl(\theta\bigg[\begin{matrix}\alpha\\ \beta\end{matrix}\bigg]\biggr)^n-
 \frac{i\pi}{4\pi^2} z_{12}^2\sum_{a,b=1}^n f(a-b)\del_{z_a}\del_{z_b}
 \prod_{c=1}^n\theta\bigg[\begin{matrix}\alpha\\ \beta\end{matrix}\bigg](z_c|\tau)\Bigg|_{z_c=0} \,.
  \end{split}
\ee
The correction term can be written:
\be
- \frac{i\pi n^2}{4\pi^2} Z^2\bigg\{\sum_{a\ne b=1}^n f(a-b)\del_{z_a}\del_{z_b}
+ f(0)\sum_{a=1}^n \del_{z_a}^2\bigg\}
 \prod_{c=1}^n\theta\bigg[\begin{matrix}\alpha\\ \beta\end{matrix}\bigg](z_c|\tau)\Bigg|_{z_c=0} \,,
\ee
from which one finds:
\be
\chi_{g} \= \Bigg(1- \frac{i\pi n^2}{4\pi^2} Z^2\bigg\{\sum_{a\ne b=1}^n f(a-b)\biggl(\frac{\theta'}{\theta}
 \bigg[\begin{matrix}\alpha\\ \beta\end{matrix}\bigg](0|\tau)\biggr)^2
 + n f(0)\frac{\theta''}{\theta} \bigg[\begin{matrix}\alpha\\ \beta\end{matrix}\bigg](0|\tau)\bigg\}
\Bigg)
\biggl(\theta\bigg[\begin{matrix}\alpha\\ \beta\end{matrix}\bigg](0|\tau)\biggr)^n \,.
\label{genusncorrterm}
\ee
It is easily shown that:
\be
\begin{split}
f(0) & \=\frac{i\pi}{6}\Bigl(1-\frac{1}{n^2}\Bigr) \,,\\
\sum_{a\ne b}f(a-b) & \=  -\frac{i\pi}{6}\Bigl(n-\frac{1}{n}\Bigr) \,,
\end{split}
\ee
where the second equation follows from $\sum_{a}f(a-b)=0$ combined with the first equation.

Inserting these into \eref{genusncorrterm}, we find the correction factor to be:
\be
1+Z^2\,\frac{n(n^2-1)}{24}\biggl(\frac{\theta''}{\theta}-\Big(\frac{\theta'}{\theta}\Big)^2\biggr) \,,
\ee
in perfect agreement with \eref{genusonecorr}.

\section{Verifying the equivalence for $n=2$ \label{sec:nequals2}}

In this section we focus on the case~$n=2$ and aim to establish an equality~$\chi_{t}=\chi_{g}$ 
in a power-series expansion in~$Z=z_{12}/2$. We do this by expanding both quantities as a power series 
in~$Z$, and expressing each coefficient as a function of~$\t$ only. The coefficients turn out to be 
functions of the Jacobi theta constants~$\theta\biggl[\begin{matrix}\alpha\\ \beta\end{matrix}\biggr](0|\tau)$,  
their derivatives, and the Eisenstein series:
\be
G_{2m}(\tau) \= \sum_{m,n\in\IZ \atop(m,n)\ne(0,0)}\frac{1}{(m\tau+n)^{2m}} \,.
\ee

The twist field expression~$\chi_{t}$ can be written for even spin structures as the product of two theta functions:
\be
\chi_{t}(Z,\t;\a,\b) 
\=\biggl( \theta\biggl[\begin{matrix}\alpha\\ \beta\end{matrix}\biggr]\Bigl(\frac{Z}{2}\Big|\tau\Bigr) \biggr)^2 \,. 
\ee
The odd spin structure~$\bigg[\begin{matrix}\half \\ \half \end{matrix}\bigg]$ has a very similar expression but has
a minus sign in various expressions compared to the even ones. We only show the intermediate steps for the 
even spin structures below, but the final result of equality of~$\chi_{t}$ and~$\chi_{g}$ holds for both even and odd 
spin structures.

We now write this as a power series (suppressing~$\a,\b$ on both sides, and denoting by~$f^{(n)}(z)$ 
the~\nth derivative with respect to~$z$):
\be
\begin{split}
\chi_{t,n=2}(Z,\tau) 
&\= \sum_{j=0}^\infty\frac{1}{(2j)!}\Big(\frac{Z}{2}\Big)^{2j}(\theta^2)^{(2j)} \,, \\
&\=\theta^2+\frac{Z^2}{8}(\theta^2)^{(2)}+\frac{Z^4}{2^4. 4!}(\theta^2)^{(4)}+
\cdots \,.
\label{Rexp}
\end{split}
\ee

Now we us consider the higher-genus expression~$\chi_{g}$ for $n=2$. In this case there are two cut 
differentials~$\w_{0}=1$ and~$\w_{1}$ given by \eref{cutdiff}. Since there is only one non-trivial cut differential,
we denote it by (with~$\theta_{1}$ the odd Jacobi theta function):
\be \label{defw2}
\w(z,Z) \= \omega_1(z,2Z)\=\frac{\theta_1(z|\tau)}{\sqrt{\theta_1(z-Z|\tau)\,\theta_1(z+Z|\tau)}} \,.
\ee
Correspondingly, we define:
\bea
A(Z) & \= & A_{01}(2z_{12}) \= \int_0^1\omega(z,Z)\,dz\,, \cr
B(Z) & \= & B_{01}(2z_{12}) \=\int_0^\tau\omega(z,Z)\,dz \,.
\label{ABdeftwo}
\eea
The period matrix~\eqref{omegadef} is given by:
\be
 \Omega=\half \begin{pmatrix} \tau+C & \tau-C\\ 
\tau-C & \tau+C\end{pmatrix} \,,
\ee
where~$C= B/A$. Defining~$\wh B$ and~$\wh C$ via the relations
\be \label{defBhat}
 \frac{B}{A} \= \tau+2\pi i\frac{\wh{B}}{A} \= \tau+2\pi i\, \wh{C}(Z) \,,
\ee
we write the period matrix as:
\be
\Omega(Z,\tau)\=\begin{pmatrix} \tau & ~0  \\ 
0 & ~\tau\end{pmatrix}+
i\pi\, \hC(Z)\begin{pmatrix} \phantom{-}1 & -1\\ 
-1 & \phantom{-}1\end{pmatrix} \,.
\ee

The higher-genus expression is given by:
\be
\chi_{g,n=2}(Z,\tau;\a,\b)
\; := \;
\frac{\Theta\bigg[\begin{matrix}\valpha_\text{diag}\\ \vbeta_\text{diag}\end{matrix}\bigg](0|\Omega)}{\sqrt{A}} \,,
\label{Lntwo}
\ee
which we want to expand in powers of~$Z$. The expansion only contains even powers of~$Z$ 
as all functions are even functions of~$Z$. 
First we expand~$\log\omega$:
\be
\log\omega(z,Z)\=-\sum_{j=1}^\infty\frac{1}{(2j)!}\bigl(\log \theta_1(z)\bigr)^{(2j)} \, Z^{2j} \,.
\ee
Expressing the logarithmic derivative of~$\theta_{1}$ in terms of the Weierstrass~$\wp$-function:
\be
\bigl(\log\theta_1(z)\bigr)''\=-\wp(z)-G_2(\tau) \,,
\label{zetap}
\ee
and using the fact that~$\w(0,Z)=1$, we obtain the expansion of the cut-differential:
\be \label{wexp}
\omega(z,Z)\=\exp\Biggl(\half G_2 \, Z^2+\sum_{j=0}^\infty\frac{\wp^{(2j)}(z)}{(2j+2)!} \, Z^{2j+2}\Biggr) \,.
\ee

Next we define the coefficients of the periods defined in~\eqref{ABdeftwo} and~\eqref{defBhat}:
\be
A(Z)\=\sum_{n=0}^\infty A_{2n} \, Z^{2n} \, , \qquad \wh B(Z) \=\sum_{n=0}^\infty B_{2n} \, Z^{2n} \,.
\ee
In order to compute these coefficients as a power series, we begin by writing the Laurent expansion around~0 of 
the Weierstrass~$\wp$-function:
\be \label{wpexp}
\begin{split}
\wp(z)&\=\frac{1}{z^2}+\sum_{m=1}^\infty (2m+1)G_{2m+2}(\tau) \, z^{2m}  \\
&\=\frac{1}{z^2}+3G_4\, z^2+5G_6\, z^4+\cdots \,,\\
\end{split}
\ee
This expansion implies the following useful equation for every integer $j>0$:
\be \label{wpderiv}
\wp^{(2j)}(z) \=(2j+1)! \, \Bigl(\frac{1}{z^{2j+2}}+G_{2j+2}\Bigr)+ O(z) \,, 
%\Bigl[\hbox{vanishing as }z\to 0\Bigr] \,.
\ee
using which we obtain the periods:
\be
\int_0^1 \wp^{(2n)}(z) \, dz\=-G_2(\tau)\delta_{n,0}\, ,\qquad
\int_0^\tau \wp^{(2n)}(z) \, dz\=\big(-\tau G_2(\tau)+2\pi i\big)\delta_{n,0} \,.
\ee

Putting this together with the expansion~\eqref{wexp}, we obtain the coefficients~$A_{2n}$, $\wh B_{2n}$.
The first few coefficients in the expansion of $A$ are:
\be
\begin{split}
A_{0} & \=  1 \, , \\
A_{2} & \=  0 \, , \\
A_{4} & \=  -\frac18 G_2^2 + \frac58 \, G_{4} \, , \\
A_{6} & \= -\frac1{24} G_2^3 - \frac38 G_4 \, G_2 + \frac{49}{24} \, G_6 \, , \\
A_{8} & \=  -\frac1{128} \, G_2^4 - \frac{17}{64} \, G_4 \, G_2^2 - \frac{25}{32} \, G_6 \, G_2 + \frac{2365}{896} \, G_4^2 \, ,
\end{split}
\ee
and those of $\wh B$ are:
\be
\begin{split}
\hB_{0} & \=  0 \, ,\\
\hB_{2} & \=  \frac12 \, ,\\
\hB_{4} & \=    \frac14 \,G_2  \,, \\
\hB_{6} & \=   \frac1{16}\,G_2^2   +\frac{11}{16} \,G_4  \,, \\
\hB_{8}  & \=    \frac1{96}  \,G_2^3 + \frac{11}{32}  \,G_4 \,G_2 +  \frac{173}{96} \,G_6 \, .
\end{split}
\ee
Finally, expanding~$\wh C$ 
as~$\hC(Z)=\sum_0^\infty \hC_{2n}Z^{2n}$, we have the first few coefficients:
\be
\begin{split}
\hC_{0} & \= 0 \, , \\
\hC_{2} & \=  \frac12\, , \\
\hC_{4} &  \=  \frac14\,G_2\, , \\
\hC_{6} & \=  \frac18 \,G_2^2 + \frac38 \,G_4  \, , \\ 
\hC_{8} & \= \frac1{16}\,G_2^3 + \frac38\,G_4\,G_2 + \frac{25}{32}\,G_6 \, , \\ 
\hC_{10} & \= \frac1{32}\,G_2^4 + \frac{9}{32} \,G_4\,G_2^2 + \frac{25}{32} \,G_6\,G_2 + \frac{9}{8} \,G_4^2  \, .
\end{split}
\label{ccoeffs}
\ee

The Siegel $\Theta$ function in~\eref{Lntwo} can be written as:
\bea
\Theta\bigg[\begin{matrix}\valpha_\text{diag}\\ \vbeta_\text{diag}\end{matrix}\bigg]
\bigl(0|\Omega(Z,\tau)\bigr)&\=& \sum_{m_1,m_2\in\IZ}\exp\sum_{a=1}^2
\Big\{ i\pi (m_a+\alpha)^2\tau+2\pi i (m_a+\alpha)\beta\Big\}\times \nn \\
&& \qquad\qquad\qquad\qquad\qquad\qquad\qquad \exp\left\{-\pi^2  (m_1-m_2)^2 \hC(Z)\right\}\,, \nn \\
&\=& \exp\left\{\sfrac14 \hC(Z) (\del_{w_1}-\del_{w_2})^2\right\}
\biggl(\theta\bigg[\begin{matrix}\alpha\\ \beta\end{matrix}\bigg](w_1|\tau)~
\theta\bigg[\begin{matrix}\alpha\\ \beta\end{matrix}\bigg](w_2|\tau)\biggr)\Bigg|_{w_i=0}\,,\nn \\
&\=& \exp\Bigl(\sfrac14 \hC(Z) \del_v^2\Bigr)
\, \theta(v|\tau)^2\Big|_{v=0} \,,
\label{Thform}
\eea
where in the last step we defined $u=\half(w_1+w_2),v=\half(w_1-w_2)$ and used the fact that we 
are working with even spin structures.
To obtain the required power-series, we expand the exponential of the operator above:
\be
\begin{split}
\exp\Bigl(\sfrac14 \hC(Z) \del_v^2\Bigr) \, 
\theta(y|\tau)^2\Big|_{v=0}
&=\theta^2 + \frac18 Z^2 (\theta^2)^{(2)}+\frac{1}{128}Z^4\Bigl(8G_2 (\theta^2)^{(2)} + (\theta^2)^{(4)}\Bigr)+ \cdots \,,
\end{split}
\ee
where we have again dropped the spin structures and arguments of the $\theta$ functions to simplify the notation.

The denominator factor in \eref{Lntwo} can also be expanded:
\be
\frac{1}{\sqrt{A(Z)}}\=\biggl(1+\Bigl(-\frac{1}{8}G_2^2+\frac58 G_4\Bigr)Z^4+\cdots\biggr)^{-\half}\\
= \; 1+ \frac{1}{16} (G_2^2 - 5 G_4) Z^4+\cdots \,.
\ee
Putting things together, we find:
\be
\chi_{g,n=2}(Z)\=\theta^2 + \frac18 Z^2 (\theta^2)^{(2)}  + \frac{1}{128}Z^4 \Bigl(8 (G_2^2-5G_4)\theta^2
+ 8G_2 (\theta^2)^{(2)} + (\theta^2)^{(4)}\Bigr) +\cdots \,.
\label{Lexpone}
\ee
For the~$O(z^{4})$ term to agree with that of~\eref{Rexp} we need to show that:
\be
8 (G_2^2-5G_4)\theta^2 + 8G_2 (\theta^2)^{(2)}\=-\frac23 (\theta^2)^{(4)} \,.
\label{zfourid}
\ee

To do so we start by writing the identity:
\be
\wp^2 -\frac16 \wp''\=5G_4 \,,
\label{psqeqn}
\ee
which follows from the periodicity property of the Weierstrass~$\wp$-function and using its Laurent 
expansion~\eqref{wpderiv} to make a linear combination regular at the origin.
(It also follows by differentiating the famous equation $(\wp')^2=4\wp^3-60 G_4 \wp-140 G_6$.) 
We remark that similar equations can be found in this manner for higher powers of $\wp$ and for 
products of derivatives, for example:
\be
\begin{split}
\wp^3 &\= \frac{1}{120}\wp'''' +9\,G_4\wp+14 \,G_6 \,, \\
\wp\wp'' &\=\frac{1}{20}\wp'''' +24\,G_4\,\wp+24\,G_6 \,.
\end{split}
\ee
Now, evaluating Eqns.(\ref{zetap}) and (\ref{psqeqn}) for $z$ successively equal 
to~$0,\half, \frac{\tau}{2},\frac{1+\tau}{2}$, 
one gets the following identity valid for all spin structures:
\be
\biggl(\Bigl(\frac{\theta'}{\theta}\Bigr)'\biggr)^2+(G_2^2-5G_4)+2G_2 \Bigl(\frac{\theta'}{\theta}\Bigr)'
\=-\frac16 \Bigl(\frac{\theta'}{\theta}\Bigr)''' \,.
\ee
Multiplying by $8\theta^2$ and rearranging the various terms, we obtain~\eref{zfourid}.

We proceed systematically in this fashion. 
Writing a power series expansion in~$Z$ for~$\chi_{g,n=2}(Z) - \chi_{t,n=2}(Z)$ and demanding that it vanishes, 
we obtain an expression at each order in~$Z$ that should identically vanish. 
The first few proposed identities are:
\bea
O(Z^{2}) &\, : \, & 0 \cr
O(Z^{4}) &\, : \, & \biggl( \frac1{16}G_2^2 + \frac1{16}G_2 \,D_{z}^2 
-\frac{5}{16}G_4 + \frac{1}{192}D_{z}^4 \biggr) \, \theta(z|\t)^{2} \Big{|}_{z=0} \cr
O(Z^{6}) &\, : \, & \biggl( \frac{1}{48}G_2^3 + \frac{5}{128}\,G_2^2 \,D_{z}^2 + G_2 \Bigl(\frac{3}{16}\,G_4 
+ \frac{1}{128}\,D_{z}^4 \Bigr)  \cr
&& \qquad \qquad \qquad \qquad + \frac{7}{128} \, G_4 \, D_{z}^2 -\frac{49}{48}\,G_6 + \frac{7}{23040}\,
D_{z}^6 \biggr) \, \theta(z| \t)^{2} \Big{|}_{z=0}  \\
O(Z^{8}) &\, : \, & \biggl(\frac{5}{512}\,G_2^4 + \frac{17}{768}\,G_2^3 \,D_{z}^2 + G_2^2 \Bigl(\frac{19}{256}\,G_4 
+ \frac{13}{2048}\,D_{z}^4 \Bigr) \cr 
&& \qquad + G_2\Bigl(\frac{25}{256}\,G_4 \,D_{z}^2 + \frac{25}{64}\,G_6 + \frac{1}{2048}\,D_{z}^6 \Bigr) \+ \cr
&& \qquad \qquad \frac{55}{512}\,G_4^2 + \frac{19}{2048}\, G_4 \,D_{z}^4 + \frac{13}{192}\,G_6 \,D_{z}^2 
-\frac{765}{256}\,G_8 + \frac{13}{1290240}\,D_{z}^8  \biggr) \, \theta(z|\t)^{2} \Big{|}_{z=0} \nn \,.
%O(Z^{8}) &\, : \, &17/3840*G2^5 + 151/12288*D*G2^4 + (5/96*G4 + 7/1536*D^2)*G2^3 + (179/2048*D*G4 + (101/768*G6 + 25/49152*D^3))*G2^2 + (15/256*G4^2 + 17/1024*D^2*G4 + (277/1536*D*G6 + (133/128*G8 + 1/49152*D^4)))*G2 + (175/4096*D*G4^2 + (245/768*G6 + 31/49152*D^3)*G4 + (101/6144*D^2*G6 + (299/2048*D*G8 + (-5687/640*G10 + 59/232243200*D^5)))) \,. 
\eea
Each expression here is built out of the Jacobi theta functions~$\theta(\tau)$,  the derivative 
operator~$D_{z}:=\frac{1}{2 i} \p_{z}$,
and the Eisenstein series~$G_{2k}(\t)$, $k=1,2,\cdots$, i.e.~they are quasi-modular forms on a
congruent subgroup of~$SL_{2}(\IZ)$ of weight~$2k$ for the expression at~$O(Z^{2k})$ (see 
e.g.~\cite{DMZ}). 
In this paper we do not give a systematic formal proof for the validity of each of these identities,
but perform a computational check of these identities. 
A proof can be constructed by using the fact that the ring of quasi-modular forms is finitely 
generated. It is therefore enough to check a finite number of coefficients in the~$q$-expansion in 
order to prove these identities. The exact number of coefficients depends on the dimension of 
the space of quasi-modular forms, and a proof can be constructed for each~$k$ by using the dimension 
formula\footnote{The dimension is typically linear in~$k$ (e.g.~the space of modular forms on~$SL(2,\IZ)$
has dimension~$k/12$ up to order one corrections), and our computations should cover these quite easily.}. 
Using the PARI/GP program~\cite{PARI}, we checked that the coefficients of the functions appearing up 
to~$O(Z^{40})$ each vanish up to~$O(q^{400})$. We consider this convincing evidence that~$\chi_{t,n=2}$ 
and~$\chi_{g,n=2}$ are equal at each order in $Z$. 

The physical intuition behind our proposal, as well as our modular forms calculations, suggest that there is a 
more formal and elegant mathematical proof of these relations. We note that relations between genus two and 
genus one theta-functions of a similar spirit, but with different physical and mathematical details, were 
proved in~\cite{Mason:2007ph, Tuite:2010yd, Tuite:2013bta}. We postpone such investigations to the future.

\section{Summing over spin structures and the thermal entropy relation \label{sec:sumspin}}

In this section we consider the sum over all spin structures of the higher-genus result, \eref{ZFullDirMaj}.
As is well-known, there are $2^{2n}$ spin structures and one is expected to sum over all of them.
There are two immediate consequences of doing so.
One is that the answer manifestly satisfies Bose-Fermi equivalence, since this is also the free boson answer
at~$R=1$. The second is that it is modular covariant, as was shown in~\cite{Lokhande:2015zma}. It only remains to
demonstrate that the result satisfies the thermal entropy relation, which, in its strong form, is actually a pair of relations 
valid respectively as $z_{12}\to 0$ and $z_{12}\to 1$. Recall that it was shown in~\cite{Lokhande:2015zma} that $\chi_t$ 
apparently cannot, in any reasonable way, be made to satisfy these relations.

For a single Majorana fermion, the sum is proportional to:
\be
\frac{1}{2^{n}}\sum_{\valpha,\vbeta}\Bigg|\Theta\biggl[\begin{matrix}\valpha \\ \vbeta\end{matrix}\biggr](0|\Omega)\Bigg| \,,
\ee
where $\valpha,\vbeta$ range over all the $2^{2n}$ spin structures on the higher-genus surface\footnote{For a Dirac fermion, everything is squared and the argument works similarly.}. Recall that~$\Omega$ 
is given in terms of $C_k$ by \eref{omegadef}. It is convenient to parametrise $C_k$ as follows. We have already seen 
that~$C_0=\tau$. As was done previously for genus 2, let us write $C_k=\tau+2\pi i \hC_k$ for $k=1,2\cdots,n-1$. 
Inserting this and using:
\be
\sum_{k=0}^{n-1}\alpha^{(a-b)k}\=n\,\delta_{ab} \,,
\ee
we get:
\be
\begin{split}
\Theta\biggl[\begin{matrix}\valpha \\ \vbeta\end{matrix}\biggr](0|\Omega)&\=\sum_{{\vec m}\in \IZ^n}\exp\left(i\pi\tau\sum_{a=1}^{n}(m_a+\alpha_a)^2 +2\pi i \sum_{a=1}^n(m_a+\alpha_a)\beta_a\right)\\
&\qquad \qquad\qquad\qquad \times
\exp\left(-\frac{2\pi^2}{n} \sum_{k=1}^{n-1}\hC_{k}\left|\sum_{a=1}^{n}\alpha^{ak}(m_a+\alpha_a)\right|^2\right) \,.
\end{split}
\label{thetaexpl}
\ee

As the interval size becomes small we have $z_{12}\to 0$ and $\hC_k\to 0$ for all $k$. In this limit, the second 
exponential tends to 1 and:
\be
\Theta\biggl[\begin{matrix}\valpha \\ \vbeta\end{matrix}\biggr](0|\Omega) \; \to \; \prod_{a=1}^n \biggl(\theta
\biggl[\begin{matrix}\alpha_a \\ \beta_a\end{matrix}\biggr](0|\tau)\biggr) \,.
\ee
It follows that:
\be
\frac{1}{2^{n}}\sum_{\valpha,\vbeta}\Bigg|\Theta\biggl[\begin{matrix}\valpha \\ \vbeta\end{matrix}\biggr](0|\Omega)\Bigg| \; \to \;  
\Biggl(\half\sum_{\alpha,\beta}\Bigg|\theta
\biggl[\begin{matrix}\alpha \\ \beta\end{matrix}\biggr](0|\tau)\Bigg|\Biggr)^n \,,
\ee
which immediately implies the small-interval part of the thermal entropy relation.

On the other hand, as the interval grows large, $z_{12}=1-\epsilon$ with $\epsilon\to 0$, we have 
$\hC_k\to  \frac{1}{\pi^2}\sin\frac{\pi k}{n}|\log\epsilon|$. Then the second term will be exponentially damped 
unless the coefficient of $|\log \epsilon|$ is zero. This requires:
\be
\sum_{k=1}^{n-1}\left|\sum_{a=1}^{n}\alpha^{ak}(m_a+\alpha_a)\right|^2\sin\frac{\pi k}{n} =0 \,.
\label{largeint}
\ee
The terms are all positive and can only vanish if each term vanishes:
\be
\sum_{a=1}^{n}\alpha^{ak}(m_a+\alpha_a)=0,\quad k=1,2,\cdots, n-1 \,.
\ee
Suppose $n$ is odd. Then due to the symmetry under $k\to n-k$, only the first $\frac{n-1}{2}$ equations are 
independent and they imply that all the $m_a+\alpha_a$ are equal, which in turn can only happen if:
\be
m_a=m,~\hbox{all }a,\qquad \alpha_a=\alpha,~\hbox{all }a \,.
\label{alphaequal}
\ee
Now, the entire dependence on the $\beta_a$ spin structure comes from the term:
\be
\exp\biggl(2\pi i \sum_{a=1}^n (m_a+\alpha_a)\beta_a\biggr) \,.
\ee
In view of \eref{alphaequal}, this can be written:
\be
\exp\biggl(2\pi i (m+\alpha)\sum_{a=1}^n \beta_a\biggr) \,.
\ee
Now each $\beta_a$ is independently equal to $0$ or $\half$ mod 1. Of the $2^n$ total choices, half of them
have~$\sum_a \beta_a=0 \, \text{(mod 1)}$ and the other half have $\sum_a\beta_a=\half \text{(mod 1)}$. 
This means that in this limit we can write:
\be
\begin{split}
\frac{1}{2^n}\sum_{\valpha,\vbeta}\bigg|\Theta\biggl[\begin{matrix}\valpha \\ \vbeta\end{matrix}\biggr]
(0|\Omega)\bigg|~ \; \to \; &~ \sum_{\alpha,\beta}\frac{2^{n-1}}{2^n} \, 
\bigg|\sum_{m\in \IZ}e^{i\pi n\tau (m+\alpha)^2+2\pi i m\beta}\bigg| \,,\\
& \= \half \sum_{\alpha,\beta}\bigg|\theta\biggl[\begin{matrix}\alpha \\ \beta\end{matrix}\biggr](0|n\tau)\bigg| \,,\\
\end{split}
\ee
from which the $z_{12}\to 1$ limit of the thermal entropy relation follows.

Notice that in the small interval limit the replica partition function goes over to the ``uncorrelated'' sum over spin structures, 
while in the large-interval limit it goes to the ``correlated'' sum\footnote{The latter statement follows by reverse applications 
of Equations~(3.5) and (3.4) in~\cite{Lokhande:2015zma} (one has to make the obvious change from the Dirac fermion case 
studied there to the Majorana fermion case above by taking a square root).}. This was exactly the behaviour argued 
in~\cite{Lokhande:2015zma} to satisfy the thermal entropy relation. Here the essential point is that we did not put it in by hand, 
rather it emerged as a property of the spin-structure-summed higher-genus $\Theta$-function.

\section{Conclusions and Outlook \label{sec:summary}}

Let us first summarise the part of our result that makes no reference to modular invariance. Suppose one wants to calculate 
the~$n$'th \Renyi entropy of free fermions on a circle at finite temperature, with fixed fermion boundary conditions around the 
space and imaginary-time axes (the former is up to us, while the latter should be anti-periodic). Then the twist-operator method 
of~\cite{Azeyanagi:2007bj} provides an answer in terms of Jacobi~$\theta$-functions, while the partition function on the 
genus-$n$ replica Riemann surface provides another answer in terms of Siegel~$\Theta$-functions. We have stated a precise 
identity, \eref{question}, which, if true, implies the equivalence of these two answers. We have provided some evidence for this 
identity for arbitrary $n$, and stronger evidence for $n=2$. One could even turn things around and argue that free fermion theory 
provides the rationale, or ``physics proof'', of our identity\footnote{We thank Edward Witten for this observation.}. Nonetheless it 
should be possible to work out a rigorous mathematical proof.

Now if we want to compute the $n$th \Renyi entropy of a {\em modular-invariant} CFT---for example the Ising model---using 
the free fermion description, then it is clear that a sum over spin structures (fermion boundary conditions) is 
required~\cite{Headrick:2012fk,Lokhande:2015zma}. Performing such a summation on the twist-operator computation 
of~\cite{Azeyanagi:2007bj} does not provide a consistent answer compatible with physical requirements like Bose-Fermi 
equivalence and the thermal entropy relation. By contrast, if we sum the genus-$n$ replica partition function over all spin 
structures in genus-$n$, we do get a consistent answer and we claim this is the correct answer for the \Renyi entropy of 
the system. This does not exclude the possibility that some twist-operator computations, so far not performed, could generalise 
that of~\cite{Azeyanagi:2007bj} to provide the complete and correct answer without recourse to the higher-genus replica partition 
function\footnote{We thank Matthias Gaberdiel and Shiraz Minwalla for this suggestion.}. Of course any proposal for such a computation must agree with the higher-genus replica partition function for each choice 
of spin structure. That would require a new set of identities generalising \eref{question} away from the diagonal replica spin 
structure.

On the way, we showed that the \Renyi entropy at finite size and temperature, after removing a universal factor,  transforms like 
a weak Jacobi form whose weight and index we obtained. In particular it is not periodic in the size of the interval or the inverse
temperature. This is to be expected, since otherwise the difference between the large-interval and small-interval entanglement 
would be zero, contradicting the thermal entropy relation. In fact we demonstrated a periodicity under $n$-fold multiples of the 
basic shifts. This means that our answer contains not just the \Renyi entropy but also its analytic continuation to a region where  
the ``entangling interval'' wraps the basic torus one or more times. It would be interesting to understand the physical meaning of 
this more general quantity.

Finally, one may hope that our understanding of the higher-genus replica surface paves the way for the study of \Renyi and 
entanglement entropies for other 2d conformal field theories at finite size and temperature (in this context see 
also~\cite{Schnitzer:2015ira,Schnitzer:2016xaj}). It may also be useful for the study of other interesting entanglement measures 
such as entanglement 
negativity~\cite{Vidal:2002zz,Calabrese:2012ew,Calabrese:2014yza, Coser:2015eba, Coser:2015dvp, Herzog:2016ohd} 
for such systems.

\section*{Acknowledgements}

We would like to thank Alejandra Castro, Bin Chen, Jan de Boer, Mathias Gaberdiel, Rajesh Gupta, Chris Herzog, 
Dileep Jatkar, Sagar Lokhande, Alex 
Maloney, Henry Maxfield, Shiraz Minwalla, Greg Moore, Mukund Rangamani, Shu-Heng Shao, Nicholas Shepherd-Barron, Tadashi 
Takayanagi, Erik Tonni, Erik Verlinde, Herman Verlinde, Edward Witten, and Ida Zadeh for interesting and useful discussions. The 
work of Sunil Mukhi was supported by a J.C.~Bose Fellowship, Government of India, that of Sameer Murthy was supported by 
the ERC Consolidator Grant N.~681908, ``Quantum black holes: A macroscopic window into the microstructure of gravity'', and that of 
Jie-qiang Wu was supported by NSFC Grant No.~11275010, No.~11335012, and No.~11325522. 
Sunil Mukhi and Jie-Qiang Wu are grateful for the warm hospitality of the Yukawa Institute of Theoretical Physics,
Kyoto where part of this work was carried out.
Sameer Murthy thanks the Aspen Center for Physics, which is supported by National Science Foundation grant
PHY-1607611, for hospitality when part of this work was performed.
Sameer Murthy and Jie-qiang Wu are grateful for the warm hospitality of IISER, Pune where part of this work was performed.

\appendix

\section{Periodicities of $\chi_{g}$ and $\chi_{t}$ \label{sec:periodicity}}

In this appendix we compute the periodicities of the expressions $\chi_{g}(z_{12},\tau;\alpha,\beta)$
and $\chi_{t}(z_{12},\tau;\alpha,\beta)$ defined in Eqs.(\ref{Ldef}) and (\ref{Rdef}) respectively.
In fact neither side is periodic under the ``simple'' translations $z_{12}\to z_{12}+1,z_{12}\to z_{12}+\tau$.
Rather, when $n$ is odd we show that the two sides are perioic under an $n$-fold shift $z_{12}\to z_{12}+n$, and quasi-periodic under the other $n$-fold shift $z_{12}\to z_{12}+n\tau$. We will find that the expressions turn out to be Jacobi forms in the  variable $Z=\frac{z_{12}}{n}$, which is (quasi)-periodic under the standard translations $Z\to Z+1,Z+\tau$. For even $n$, things are slightly different. In this case the transformations $z_{12}\to z_{12}+n,z_{12}\to z_{12}+n\tau$ lead to a change in spin structures for both $\chi_g$ and $\chi_t$, in addition to a pre-factor in the latter case. The change in spin structures, as well as the pre-factors, are the same on both sides. For genuine (quasi)-periodicity at even $n$, one has to further double the shifts and consider $z_{12}\to z_{12}+2n,z_{12}+2n\tau$ and we find that $\chi_g$ and $\chi_t$ transform in the same way under these transformations.

\subsection{Higher-genus calculation}

For $\chi_{g}$, we start by examining the effect of the shifts $z_{12}\to z_{12}+n,z_{12}\to z_{12}+n\tau$
on $C_k=\frac{B_{0k}}{A_{0k}}$ where
$B_{0k}=\int_{b_0}\omega_k$ and $A_{0k}=\int_{a_0}\omega_k$. The cut differentials depend on
$\theta_1 \equiv \theta\bigg[\begin{matrix}\half \\ \half\end{matrix}\bigg]$.
We use the fact that~\cite{Mumford:book}:
\be
\begin{split}
\theta\bigg[\begin{matrix}\alpha\\ \beta\end{matrix}\bigg]
(z+n|\tau)&\=e^{2\pi i \alpha n}\,\theta\bigg[\begin{matrix}\alpha\\ \beta\end{matrix}\bigg]
(z| \tau) \,, \\
\theta\bigg[\begin{matrix}\alpha\\ \beta\end{matrix}\bigg](z+n\tau| \tau)
&\=e^{-2\pi i \beta n}e^{-i\pi n^2\tau}e^{-2\pi i n z}\,\theta\bigg[\begin{matrix}\alpha\\ \beta\end{matrix}\bigg](z| \tau) \,.
\end{split}
\label{thetaper}
\ee
Inserting this for $\bigg[\begin{matrix}\alpha \\ \beta\end{matrix}\bigg]=\bigg[\begin{matrix}\half \\ \half \end{matrix}\bigg]$
 in \eref{cutdiff} we find that the cut differentials transform as:
\be
\begin{split}
\omega_k (z,z_{12}+n,\tau)&\=\omega_k (z,z_{12},\tau) \,, \\[2mm]
\omega_k (z,z_{12}+n\tau,\tau)&\=e^{2\pi i(n-k)\frac{k}{n}z_{12}}e^{\pi i\tau k(n-k)}\omega_k(z,z_{12},\tau) \,.
\end{split}
\ee
For the integrals $A_{0k},B_{0k}$, these shifts in $z$ have the effect of deforming the contour of integration. We can represent the effect of this deformation as follows:
\be
\begin{split}
z_{12}\to z_{12}+n:\qquad
&\int_A\omega_k\rightarrow \int_A\omega_k,\qquad \int_B\omega_k\rightarrow \int_{B}\omega_k\pm\, n\!\!\int_{A}\omega_k \,, \\[4mm]
z_{12}\rightarrow z_{12}+n\tau:\qquad
&\int_{A}\omega_k\rightarrow \int_{A}\omega_k \pm\, n \int_{B}\omega_k,\qquad
\int_{B}\omega_k\rightarrow \int_{B}\omega_k \,.
\end{split}
\ee
The sign depends on how to take the analytic extension. However in the exponential the final result does not depend on the sign. Thus, we have:
\be
\begin{split}
A_{0k}(z_{12}+n)&\=A_{0k}(z_{12}) \,, \\
B_{0k}(z_{12}+n)&\=\begin{cases}
B_{0k}(z_{12}) ~~~\mbox{for }k=0 \\
B_{0k}(z_{12})\pm nA_{0k}(z_{12}) ~~~\mbox{for }k \neq 0
\end{cases} \,;
\end{split}
\ee
and:
\be
\begin{split}
A_{0k}(z_{12}+n\tau)&\=
\begin{cases}
A_{0k}(z_{12})~~~\mbox{for }k=0 \\
e^{2\pi i(n-k)\frac{k}{n}z_{12}}e^{\pi i\tau k(n-k)}(A_{0k}(z_{12})-nB_{0k}) ~~~\mbox{for }k \neq 0
 \end{cases} \,;
\\
B_{0k}(z_{12}+n\tau)&\= e^{2\pi i(n-k)\frac{k}{n}z_{12}}e^{\pi i\tau k(n-k)}B_{0k}(z_{12}) \,.
 \end{split}
\ee
As a result, the period matrix transforms as:
\be
\begin{split}
\Omega(z_{12}+n)&\=\Omega(z_{12})\pm {\mathbf B} \,, \\
(\Omega^{-1})(z_{12}+n\tau)&\=(\Omega^{-1})(z_{12})\mp {\mathbf B} \,,
\end{split}
\label{pertrans}
\ee
where ${\mathbf B}$ is a symmetric matrix given by:
\be
{\mathbf B}_{ab}\=-1+n\delta_{ab} \,.
\label{Bmatrix}
\ee

The transformation property of the $\Theta$-function under these shifts is as follows.
Consider  the first case with $\Omega\to\Omega + {\mathbf B}$. Then:
\be
\begin{split}
\Theta\bigg[\begin{matrix}\valpha_\text{diag}\\ \vbeta_\text{diag}\end{matrix}\bigg](0|\Omega+{\mathbf B})
&=\sum_{\vm\in(\IZ)^n}
\exp\Big(i\pi ({\vec m}+\valpha_\text{diag})\cdot \Omega\cdot ({\vec m}+\valpha_\text{diag})
+2\pi i ({\vec m}+\valpha_\text{diag})\cdot\vbeta_\text{diag}\Big) \times \\
&\qquad\qquad\qquad \exp\Big(i\pi ({\vec m}+\valpha_\text{diag})\cdot \boldB\cdot ({\vec m}+\valpha_\text{diag})\Big)
\end{split}
\ee
Using the expression for $\boldB$ above, one can easily show that:
\be
\begin{split}
({\vec m}+\valpha_\text{diag})\cdot \boldB\cdot ({\vec m}+\valpha_\text{diag})&=
(n-1)\sum_a(m_a+\alpha)^2 -2\sum_{a<b}(m_a+\alpha)(m_b+\alpha)\\
&= (n-1)\sum_a m_a^2 -2\sum_{a<b}m_a m_b
\end{split}
\label{extraterm}
\ee
The second equality is obvious when $\alpha=0$, but it is easy to verify that it is also true when $\alpha=\half$.

Now when $n$ is odd, the last line of \eref{extraterm} is even and therefore it does not modify the $\Theta$ function. Hence for odd $n$ we have proved that:
\be
\Theta\bigg[\begin{matrix}\valpha_\text{diag}\\ \vbeta_\text{diag}\end{matrix}\bigg](0|\Omega+{\mathbf B})\=
\Theta\bigg[\begin{matrix}\valpha_\text{diag}\\ \vbeta_\text{diag}\end{matrix}\bigg](0|\Omega) \,,
\ee
It follows that:
\be
\chi_{g}(z_{12}+n,\tau;\alpha,\beta)\=\chi_{g}(z_{12},\tau;\alpha,\beta) \,.
\label{Ln}
\ee
We will return to the case of even $n$ below.

To study $z_{12}\rightarrow z_{12}+n\tau$, we perform the modular transformation $\Omega\to -\Omega^{-1}$
to re-write the $\Theta$-function as:
\be
\begin{split}
\Theta\bigg[\begin{matrix}
\valpha_\text{diag}  \\
\vbeta_\text{diag}
\end{matrix} \bigg]
\Big(0| \Omega(z_{12})\Big)
&\=\frac{1}{\det^\half\big(-i\Omega(z_{12})\big)}\Theta\bigg[\begin{matrix}
\ \vbeta_\text{diag}  \\ -\valpha_\text{diag} \end{matrix} \bigg]
\Big(0| -\Omega(z_{12})^{-1}\Big)\,,\\
&\=\prod_{k=1}^{\frac{n-1}{2}}\frac{1}{(-iC_k)}\Theta\bigg[\begin{matrix}
\ \vbeta_\text{diag}  \\ -\valpha_\text{diag}\end{matrix} \bigg]
\Big(0| -\Omega(z_{12})^{-1}\Big) \,. \\
\end{split}
\ee
Thus we can re-write the quantity $\chi_{g}$ as:
\be
\chi_{g}(z_{12},\tau;\alpha,\beta)
\=\prod_{k=1}^{\frac{n-1}{2}}\frac{1}{(-iB_{0k})}
\Theta\bigg[\begin{matrix}\ \vbeta_\text{diag}  \\ -\valpha_\text{diag}\end{matrix} \bigg] \Big(0| -\Omega(z_{12})^{-1}\Big) \,.
\ee
Now from \eref{pertrans} we see that under $z_{12}\rightarrow z_{12}+n\tau$, $\Omega^{-1}$ shifts by
the matrix $\mathbf B$ defined there. Using \eref{extraterm} we again find that the $\Theta$-function in the above equation is invariant for odd $n$. Hence the only change in $\chi_{g}$ comes from:
\be
\prod_{k=1}^{\frac{n-1}{2}}B_{0k}(z_1+n\tau,z_2)
\=e^{i\pi\frac{n(n^2-1)}{12}\tau}e^{i\pi\frac{n^2-1}{6} z_{12}}
\prod_{k=1}^{\frac{n-1}{2}}B_{0k}(z_1,z_2) \,,
\label{Bperiodicity}
\ee
and we finally get:
\be
\chi_{g}(z_{12}+n\tau,\tau;\alpha,\beta)\=
e^{-i\pi\frac{n(n^2-1)}{12}\tau}e^{-i\pi\frac{n^2-1}{6} z_{12}}
\chi_{g}(z_{12},\tau;\alpha,\beta) \,.
\label{Lntau}
\ee

Repeating the procedure for even $n$, the result is slightly different. One can verify that under $z_{12}\to z_{12}+n$ there is a change in spin structures:
\be
\bigg[\begin{matrix}
\valpha_\text{diag}  \\
\vbeta_\text{diag}
\end{matrix} \bigg] \to
\bigg[\begin{matrix}
\valpha_\text{diag}  \\
\vbeta_\text{diag}\pm\vhalf
\end{matrix} \bigg]
\ee
Hence $\chi_g$ for a {\em fixed} spin structure does not come back to itself, so for even $n$ we must consider the shift $z_{12}\to z_{12}+2n$. In this case, the matrix ${\mathbf B}$ of \eref{Bmatrix} is replaced by $2{\mathbf B}$\footnote{Naively replacing $n$ by $2n$ on the RHS of \eref{Bmatrix} is not correct, one has to re-do the derivation of this equation for the shift $z_{12}\to z_{12}+2n$ and one finds that the new ${\mathbf B}_{ab}$ is $-2+2n\delta_{ab}$.}. Using this matrix and repeating the above manipulations, one easily finds that the $\Theta$-function is invariant. Thus for even $n$ the analogue of \eref{Ln} is:
\be
\chi_{g}(z_{12}+2n,\tau;\alpha,\beta) =\chi_{g}(z_{12},\tau;\alpha,\beta) \,.
\label{Levenn}
\ee

For the other shift, $z_{12}\to z_{12}+n\tau$ again changes the spin structure, so we consider instead $z_{12}\to z_{12}+2n\tau$. With the double shift we find:
\be
\omega_k (z,z_{12}+2n\tau,\tau)\=e^{4\pi i(n-k)\frac{k}{n}z_{12}}e^{4\pi i\tau k(n-k)}\omega_k(z,z_{12},\tau) \,.
\ee
and it follows that, for even $n$, the second periodicity is:
\be
\chi_{g}(z_{12}+2n\tau,\tau;\alpha,\beta) =
e^{-i\pi\frac{n(n^2-1)}{3}\tau}e^{-i\pi\frac{n^2-1}{3} z_{12}}
\chi_{g}(z_{12},\tau;\alpha,\beta) \,.
\label{Levenntau}
\ee

\subsection{Twist operator calculation}

Next let us compute the periodicity of $\chi_{t}$ and compare. Again we start with odd $n$. In this case, the quantity $k$ appearing in $\chi_t$ is an integer. Under $z_{12}\to z_{12}+n$, the argument of the numerator $\theta$-function
shifts by this integer and we have:
\be
\begin{split}
\theta\bigg[\begin{matrix}\alpha\\ \beta\end{matrix}\bigg]\Big(\frac{k}{n}(z_{12}+n)|\tau\Big)&\=
\theta\bigg[\begin{matrix}\alpha\\ \beta\end{matrix}\bigg]\Big(\frac{k}{n}z_{12}+k|\tau\Big) \,,\\
&\=e^{2\pi i \alpha k}\, \theta\bigg[\begin{matrix}\alpha\\ \beta\end{matrix}\bigg]\Big(\frac{k}{n}z_{12}|\tau\Big) \,.
\end{split}
\ee
It follows immediately that the product over $k$ remains unchanged. Thus we have shown that for odd $n$,
\be
\chi_{t}(z_{12}+n,\tau;\alpha,\beta)\=\chi_{t}(z_{12},\tau;\alpha,\beta) \,.
\ee
which agrees with \eref{Ln}.

On the other hand under $z_{12}\to z_{12}+n\tau$, the $\theta$-function in the numerator
of $\chi_{\rm twist~field}$ has its argument shifted by $k\tau$. Thus, from \eref{thetaper}:
\be
\begin{split}
\theta\bigg[\begin{matrix}\alpha\\ \beta\end{matrix}\bigg]\Big(\frac{k}{n}(z_{12}+n\tau)|\tau\Big)&\=
\theta\bigg[\begin{matrix}\alpha\\ \beta\end{matrix}\bigg]\Big(\frac{k}{n}z_{12}+k\tau|\tau\Big) \,, \\
&\=e^{-2\pi i \beta k}e^{-i\pi k^2\tau}e^{-2\pi i \frac{k^2}{n}z_{12}}\,
\theta\bigg[\begin{matrix}\alpha\\ \beta\end{matrix}\bigg]\Big(\frac{k}{n}z_{12}|\tau\Big) \,.
\end{split}
\ee
Taking the product over $k$, we have:
\be
\begin{split}
\prod_{k=-\frac{n-1}{2}}^{\frac{n-1}{2}}\theta\bigg[\begin{matrix}\alpha\\ \beta\end{matrix}\bigg]
\Big(\frac{k}{n}(z_1+n\tau-z_2)|\tau\Big)&\=
e^{-i\pi \frac{n(n^2-1)}{12}\tau}e^{-i\pi  \frac{n^2-1}{6}z_{12}} \times \\[-5mm]
&\qquad\qquad\prod_{k=-\frac{n-1}{2}}^{\frac{n-1}{2}}\theta\bigg[\begin{matrix}\alpha\\ \beta\end{matrix}\bigg]\Big(\frac{k}{n}z_{12}|\tau\Big) \,.
\end{split}
\ee
It follows that:
\be
\chi_{t}(z_{12}+n\tau,\tau;\alpha,\beta)\=
e^{-i\pi\frac{n(n^2-1)}{12}\tau}e^{-i\pi\frac{n^2-1}{6} z_{12}}
\chi_{t}(z_{12},\tau;\alpha,\beta) \,.
\ee
This is exactly the same as the periodicity computed for $\chi_{g}$ in \eref{Lntau}.

Finally, we consider the periodicity of $\chi_t$ for even $n$. This time the quantity $k$ appearing in the argument of the $\theta$-functions is a half-integer. Hence under $z_{12}\to z_{12}+n$, the $\theta$-functions shift by half-periods and this changes their spin structure:
\be
\bigg[\begin{matrix}\alpha\\ \beta\end{matrix}\bigg]\to\bigg[\begin{matrix}\alpha\\ \beta\pm\half \end{matrix}\bigg]\
\ee
So to find periodic behaviour for a fixed spin structure, one has to consider $z_{12}\to z_{12}+2n$. It is easily verified that:
\be
\chi_{t}(z_{12}+2n,\tau;\alpha,\beta) =\chi_{t}(z_{12},\tau;\alpha,\beta) \,.
\ee

Using:
\be
\begin{split}
\theta\bigg[\begin{matrix}\alpha\\ \beta\end{matrix}\bigg]\Big(\frac{k}{n}(z_{12}+2n\tau)|\tau\Big)&\=
\theta\bigg[\begin{matrix}\alpha\\ \beta\end{matrix}\bigg]\Big(\frac{k}{n}z_{12}+2k\tau|\tau\Big) \,, \\
&\=e^{-4\pi i \beta k}e^{-4i\pi k^2\tau}e^{-4\pi i \frac{k^2}{n}z_{12}}\,
\theta\bigg[\begin{matrix}\alpha\\ \beta\end{matrix}\bigg]\Big(\frac{k}{n}z_{12}|\tau\Big) \,.
\end{split}
\ee
we easily find that, for even $n$:
\be
\chi_{t}(z_{12}+2n\tau,\tau;\alpha,\beta)\=
e^{-i\pi\frac{n(n^2-1)}{3}\tau}e^{-i\pi\frac{n^2-1}{3} z_{12}}
\chi_{t}(z_{12},\tau;\alpha,\beta) \,.
\ee
in perfect agreement with the higher-genus result in \eref{Levenntau}.

Thus we have shown that $\chi_{g}(z_{12},\tau;\alpha,\beta)$ and $\chi_{t}(z_{12},\tau;\alpha,\beta)$
have exactly the same periodicities under $z_{12}\to z_{12}+n,z_{12}\to z_{12}+n\tau$ for odd $n$, and $z_{12}\to z_{12}+2n,z_{12}+2n\tau$ for even $n$. This is a necessary criterion for the equality of the two. From this it follows that~$\chi_{\rm higher~genus}$  and~$\chi_{\rm twist~field}$, defined in Eqs.(\ref{chihg}) and (\ref{chitf}) respectively, transform the same way. In fact each acquires a pure phase, so that the
corresponding partition functions obtained by taking the modulus-squared of $\chi$ (and summing over spin
structures if necessary) are periodic---as they should be.

\section{Modular transformations of $\chi_{g}(z_{12},\tau;\alpha,\beta)$ and $\chi_{t}(z_{12},\tau;\alpha,\beta)$
\label{sec:modular}}

\subsection{Higher-genus calculation}

We start by considering the $T$ modular transformation, $\tau \rightarrow \tau+1$. Under this, one has:
\be
\int_0^{1} dz \rightarrow \int_0^1 dz \,, \qquad \int_0^{\tau} dz \rightarrow \int_0^1 dz +\int_0^{\tau} dz  \,,
\ee
and also:
\be
\begin{split}
\theta\bigg[\begin{matrix}
\alpha \\ \beta\end{matrix} \bigg](z,\tau+1)&\=e^{-\pi i\alpha(\alpha+1)} \,
\theta\bigg[\begin{matrix}\alpha \\ \alpha+\beta+\frac{1}{2}
\end{matrix} \bigg](z,\tau) \,, \\
\theta\bigg[\begin{matrix}
\alpha \\ \beta \end{matrix}\bigg]
\left(\frac{z}{\tau},-\frac{1}{\tau}\right)
&\=e^{2\pi i\alpha\beta} \,
(-i\tau)^{\frac{1}{2}} \, e^{\frac{\pi i z^2}{\tau}} \,
\theta\bigg[\begin{matrix}
\beta \\
-\alpha
\end{matrix} \bigg](z,\tau)  \,.
\end{split}
\label{thetamod}
\ee
From these it follows, recalling the definitions in Eq.(\ref{ABdef}), that:
\be
\begin{split}
A_{0k}(z_{12},\tau+1)&\=A_{0k}(z_{12},\tau) \,, \\
B_{0k}(z_{12},\tau+1)&\=B_{0k}(z_{12},\tau)+A_{0k}(z_{12},\tau) \,.
\end{split}
\ee
From this one finds that:
\be
\Omega_{jk}(z_{12},\tau+1)\=\Omega_{jk}(z_{12},\tau)+\delta_{jk}  \,.
\ee
Using the definition of the higher-genus $\Theta$-function one easily verifies that:
\be
\chi_{g}(z_{12},\tau+1;\alpha,\beta)\=
e^{-i\pi n\alpha(\alpha+1)}\chi_{t}(z_{12},\tau;\alpha,\alpha+\beta+\shalf) \,.
\label{Lmodone}
\ee

The other transformation $\tau\rightarrow -\frac{1}{\tau}$ is a little more complicated. In this case the $z$-coordinate also changes, and we have:
\be
\begin{split}
A_{0k}\Bigl(\frac{z_{12}}{\tau},-\frac{1}{\tau}\Bigr)
&\=\int_0^1 dz ~e^{\frac{\pi i}{\tau}\frac{k(k-n)}{n^2} z_{12}^2}  w_k(z\tau,z_{12},\tau) \,, \\
&\= e^{\frac{\pi i}{\tau}\frac{k(k-n)}{n^2}z_{12}^2} \int_0^{\tau} \frac{dy}{\tau} w_k(y,z_1,z_2,\tau) \,, \\
&\= e^{\frac{\pi i}{\tau}\frac{k(k-n)}{n^2}z_{12}^2} \, \frac{1}{\tau} \, B_{0k}(z_{12},\tau) \,;
\end{split}
\ee
and:
\be
\begin{split}
 B_{0k}\Bigl(\frac{z_{12}}{\tau},-\frac{1}{\tau}\Bigr) &\=
 \int_0^{-\frac{1}{\tau}} dz~ e^{\frac{\pi i}{\tau}\frac{k(k-n)}{n^2}z_{12}^2}  w_k(z\tau,z_{12},\tau) \,, \\
&\=e^{\frac{\pi i}{\tau}\frac{k(k-n)}{n^2}z_{12}^2} \int_0^{-1} \frac{dy}{\tau} w_k(y,z_{12},\tau) \,, \\
&\=-e^{\frac{\pi i}{\tau}\frac{k(k-n)}{n^2}z_{12}^2} \, \frac{1}{\tau} \, A_{0k}(z_1,z_2,\tau) \,.
\end{split}
\ee
As a consequence, we see that $C_k\to -\frac{1}{C_k}$ (recall \eref{Ckdef}) and hence, from \eref{Ominv2},
\be
\Omega_{jk}\Bigl(\frac{z_{12}}{\tau},-\frac{1}{\tau}\Bigr)\=-(\Omega^{-1})_{jk}(z_1,z_2,\tau) \,.
\ee
It then follows immediately that:
\be
\chi_{g}\Bigl(\frac{z_{12}}{\tau},-\frac{1}{\tau};
\alpha,\beta\Bigr)
\=(-i\tau)^{\frac{n}{2}}
e^{\frac{\pi i}{\tau}\, \frac{1}{12n}(n^2-1)z_{12}^2} \, e^{2\pi i n\alpha\beta} \, \chi_{t}(z_{12},\tau;\beta, -\alpha) \,.
\label{Lmodtwo}
\ee

\subsection{Twist-operator calculation}

For the twist-field calculation, we only need the properties in \eref{thetamod} of the Jacobi theta-functions.
Using these, we find that $\chi_{t}(z_{12},\tau)$ has the modular transformations:
\be
\chi_{t}(z_{12},\tau+1;\alpha,\beta)
\=e^{-i\pi n\alpha(\alpha+1)}\,\chi_{t}(z_{12},\tau;\alpha,\alpha+\beta+\shalf) \,,
\ee
and:
\be
\begin{split}
\chi_{t}\left(\frac{z_{12}}{\tau},-\frac{1}{\tau};\alpha, \beta\right)
&\=\prod_{k=-\frac{n-1}{2}}^{\frac{n-1}{2}}
\theta \bigg[ \begin{matrix}\alpha \\ \beta\end{matrix} \bigg]
\bigg(\frac{k}{n}\frac{z_{12}}{\tau}\bigg|-\frac{1}{\tau}\bigg) \,, \\
&\=\prod_{k=-\frac{n-1}{2}}^{\frac{n-1}{2}}(-i\tau)^{\frac{1}{2}}e^{\frac{\pi i}{\tau}\frac{k^2}{n^2}z_{12}^2} \,
e^{2\pi i\alpha\beta}\,
\theta \bigg[\begin{matrix}\beta \\ -\alpha \end{matrix} \bigg]
\bigg(\frac{k}{n}z_{12} \bigg| \tau\bigg) \,, \\
&\=(-i\tau)^{\frac{n}{2}}e^{\frac{\pi i}{\tau}\frac{1}{12n}(n^2-1)z_{12}^2} \, e^{2\pi i\alpha\beta n}
\, \chi_{t}(z_{12},\tau; \beta, -\alpha) \,.
\end{split}
\ee
Comparing with Eqs.(\ref{Lmodone}) and (\ref{Lmodtwo}), we see that the modular transformations of the
two sides are identical.

\bibliography{spinstr}
\bibliographystyle{JHEP}

\providecommand{\href}[2]{#2}\begingroup\raggedright\endgroup

\end{document}